\documentclass[a4paper,11pt]{article}
\pdfoutput=1

\usepackage{jheppub}
\usepackage{comment}
\usepackage{forest}
\usepackage{amsfonts}
\usepackage{cleveref}
\usepackage{srcltx}
\usepackage{bm}
\usepackage{siunitx}
\usepackage{mathdots}
\usepackage{color}
\usepackage{dsfont}
\usepackage{slashed}
\usepackage{physics}
\usepackage{subfigure}
\usepackage{tikz}
\usepackage{xspace}
\usepackage{braket}
\usepackage{orcidlink}
\usepackage{enumitem}

\newcommand{\Slash}[1]{{\ooalign{\hfil/\hfil\crcr$#1$}}}

\newcommand{\he}{$^3$He\xspace}
\newcommand{\SO}[1]{\mathrm{SO}(#1)}
\newcommand{\U}[1]{\mathrm{U}(#1)}
\newcommand{\Ao}{A$_1$\xspace}
\newcommand{\At}{A$_2$\xspace}
\newcommand{\Bt}{B$_2$\xspace}
\newcommand{\magc}{{\hat{d}}^\dagger}
\newcommand{\maga}{{\hat{d}}}
\newcommand{\cavc}{{\hat{c}}^\dagger}
\newcommand{\cava}{{\hat{c}}}
\newcommand{\domg}{\Delta\omega}

\newcommand{\taum}{\tau_\mathrm{mag}}
\newcommand{\Gamm}{\Gamma_\mathrm{mag}}
\newcommand{\geff}{g_\mathrm{eff}}
\newcommand{\tint}{T_{\mathrm{int}}}
\newcommand{\ncp}{n_{\mathrm{C}}}
\newcommand{\Ncp}{N_{\mathrm{C}}}

\DeclareSIUnit{\angstrom}{\textup{\AA}}

\title{Axion detection via superfluid $^3$He ferromagnetic phase and quantum measurement techniques}

\author[a,b]{So Chigusa\,\orcidlink{0000-0001-6005-4447},}\emailAdd{sochigusa@lbl.gov}
\author[c]{Dan Kondo,}\emailAdd{dan.kondo@ipmu.jp}
\author[a,b,c,1]{Hitoshi Murayama\,\orcidlink{0000-0001-5769-9471}\note{Hamamatsu Professor},}\emailAdd{hitoshi@berkeley.edu}
\author[c]{Risshin Okabe\,\orcidlink{0000-0002-5351-174X},}\emailAdd{risshin.okabe@ipmu.jp}
\author[d]{and Hiroyuki Sudo\,\orcidlink{0000-0003-4744-3100}}\emailAdd{h.sudo@issp.u-tokyo.ac.jp}

\affiliation[a]{Ernest Orlando Lawrence Berkeley National Laboratory, Berkeley, CA 94720, USA}
\affiliation[b]{Department of Physics, University of California, Berkeley, CA 94720, USA}
\affiliation[c]{Kavli Institute for the Physics and Mathematics of the Universe (WPI), The University of Tokyo Institutes for Advanced Study, The University of Tokyo, Kashiwa, Chiba 277-8583, Japan}
\affiliation[d]{Institute for Solid State Physics, The University of Tokyo, Kashiwa, Chiba 277-8581, Japan}

\abstract{
    We propose to use the nuclear spin excitation in the ferromagnetic A$_1$ phase of the superfluid $^3$He for the axion dark matter detection.
    This approach is striking in that it is sensitive to the axion-nucleon coupling, one of the most important features of the QCD axion introduced to solve the strong CP problem.
    We review a quantum mechanical description of the nuclear spin excitation and apply it to the estimation of the axion-induced spin excitation rate.
    We also describe a possible detection method of the spin excitation in detail and show that the combination of the squeezing of the final state with the Josephson parametric amplifier and the homodyne measurement can enhance the sensitivity.
    It turns out that this approach gives good sensitivity to the axion dark matter with the mass of $\mathcal{O}(1) \,\si{\micro eV}$ depending on the size of the external magnetic field.
    We estimate the parameters of experimental setups, e.g., the detector volume and the amplitude of squeezing, required to reach the QCD axion parameter space.
}

\begin{document}

\maketitle

\section{Introduction}

Axion \cite{Peccei:1977hh} is a proposed solution to the strong CP problem, namely to explain why the quantum chromodynamics (QCD) does not violate the time-reversal symmetry. The experimental upper limit on the neutron electric dipole moment $d_n < 1.8 \times 10^{-26}\,e\,\si{cm}$ \cite{Abel:2020pzs} implies that the so-called vacuum angle of QCD to be extremely small $\abs{\bar{\theta}} < 10^{-10}$. The theory assumes a new global $\U{1}$ Peccei--Quinn symmetry broken spontaneously at the energy scale called the axion decay constant $f_a$ as well as explicitly by the QCD anomaly. The effective operator of the axion coupling to gluons is
\begin{align}
    {\cal L}_a = \frac{g_s^2}{64\pi^2}
    \left( \bar{\theta} + \frac{a}{f_a} \right)
    \epsilon^{\mu\nu\rho\sigma} G^{b}_{\mu\nu} G^b_{\rho\sigma}\ .
\end{align}
Switching to the chiral Lagrangian, it can be shown that the axion settles to the ground state where $\bar{\theta}$ is dynamically canceled.

Interestingly, it was pointed out that the axion can also comprise the dark matter of the Universe from misalignment mechanism or emission from topological defects \cite{Preskill:1982cy,Vilenkin:1982ks}. The initial version of the theory assumed $f_a = v_\mathrm{EW}$ (electroweak scale) and was excluded by beam dump experiments~\cite{Asano:1981nh}. It was later proposed to take $f_a \gg v_\mathrm{EW}$ dubbed ``invisible axion'' \cite{Kim:1979if,Shifman:1979if,Dine:1981rt,Zhitnitsky:1980tq}. The axion abundance is higher for higher $f_a$, and $f_a \simeq \SI{e12}{GeV}$ is typically regarded as a preferred range. It translates to $m_a \simeq \si{\micro eV}$ scale.

Many direct detection experiments for the dark matter axion, such as refs.~\cite{asztalos2010squid, ADMX:2018gho, ADMX:2019uok, ADMX:2021nhd, HAYSTAC:2018rwy, HAYSTAC:2020kwv, HAYSTAC:2023cam, McAllister:2017lkb, Quiskamp:2022pks, Alesini:2019ajt, Alesini:2020vny, Alesini:2022lnp, Lee:2020cfj, Jeong:2020cwz, CAPP:2020utb, Lee:2022mnc, Kim:2022hmg, Yi:2022fmn}, rely on the axion coupling to photons $a F_{\mu\nu} \tilde{F}^{\mu\nu}$. Their prospect in the near future is becoming exciting. Yet the axion coupling to photons is highly model-dependent. To fully verify that the axion solves the strong CP problem, measuring its coupling to hadrons would be crucial. In particular, the axion couples to the nucleon spins $\vec{\nabla}a \cdot \vec{s}_N$ with relatively little model dependence. Search for dark matter axion using the nuclear spins, or confirming detected axion signal with nuclear signs, would be crucial to enhance our understanding of both the strong CP problem as well as the nature of dark matter.
In spite of its importance, there are relatively few experiments and proposals including refs.~\cite{JacksonKimball:2017elr, Wu:2019exd, Garcon:2019inh, Bloch:2021vnn, Bloch:2019lcy, Lee:2022vvb, Graham:2020kai, Gao:2022nuq, Dror:2022xpi, Brandenstein:2022eif, Wei:2023rzs, Chigusa:2023hmz} in this direction.

In this paper, we propose a new experimental technique to detect dark matter axions using their coupling to nuclear spins. Interactions among the nuclear spins are very weak because their magnetic moments are suppressed by the nucleon mass $\mu_N = e/m_N$ rather than the electron mass $\mu_B = e/m_e$. One needs to identify material where nuclear spins play a major role at very low temperatures. 

We point out that the \Ao phase of superfluid $^3$He is a unique material that has an ordering of nuclear spins without relying on their coupling to electron spins. This is because the Cooper pairs of \he atoms are in the $p$-wave (anti-symmetric) with total spin $S=1$ (symmetric) as required by Fermi statistics. In a high magnetic field, it becomes basically a ferromagnet of nuclear spins. The corresponding nuclear magnon is gapped due to the external magnetic field and the gap can be tuned to the axion mass. It is quite remarkable that the gap happens to be in the range of the preferred axion mass for dark matter with an achievable magnetic field. Then the magnon can be converted to a cavity photon resonantly due to the polariton mixing between the magnon and photon. Again the size of the cavity is such that it can be fitted in a laboratory.
Note that our setup is distinct from other proposals to use superfluid \he for axion dark matter search \cite{Gao:2022nuq, Dror:2022xpi} in the superfluid phase used and/or the signal detection method.

Because our experiments are performed at such low temperatures $T \lesssim \SI{3}{mK}$ that the target \he shows superfluidity, the quantum noise \cite{caves1982quantum} becomes non-negligible. 
These days several applications of quantum measurement techniques to axion detections have been studied in order to circumvent the quantum noise \cite{Malnou:2018dxn,HAYSTAC:2020kwv,Wurtz:2021cnm,Zheng:2016qjv,Ikeda:2021mlv,Sushkov:2023fjw,Dixit:2020ymh,Shi:2022wpf,Lamoreaux:2013koa}.
In this paper, we apply the squeezing technique, which has been discussed in refs.~\cite{Malnou:2018dxn,HAYSTAC:2020kwv}, and evaluate the improvement in the sensitivity of our experiment. 

This paper is organized as follows.
In \cref{sec:3He}, we review the properties of \he.
We analyze superfluid phases of \he using the spinor BEC formalism and understand the properties of nuclear magnons in the ferromagnetic \Ao phase.
In \cref{sec:Axiondetection}, we discuss how the axion dark matter signal can be detected using superfluid \he; we use a nuclear magnon mode, which is converted into a cavity photon through the polariton mixing.
We also discuss how noise reduction is realized by using squeezing and the homodyne measurement.
We show sensitivities for several different setups in \cref{sec:sensitivity} and conclude in \cref{sec:conclusion}.
A detailed description of our noise estimate and statistical treatment is summarized in \cref{sec:SNR}.
Finally, we review the Josephson parametric amplifier (JPA), which is a representative apparatus for squeezing, in \cref{sec:JPA}.

\section{Understanding \he via spinor BEC}\label{sec:3He}

In this section, we will describe the phase structure of the superfluid \he using Ginzburg--Landau formalism and simplified spinor BEC formalism. 
We summarize the phase structure in \cref{tab:superfluid_phases}. 
We utilize an \Ao phase for axion detection, which has a ferromagnetic property, in this paper.


\begin{table}[t]
    \centering
    \caption{Superfluid phases of \he}
    \vspace{0.3\baselineskip}
    \begin{tabular}{lll}
        \hline
        External magnetic field $H$ & Phases & Magnetic property \\ \hline
        $H=0$       & A phase & -- \\
                    & B phase & -- \\ 
        \hline
        $H\neq 0$   & \Ao phase & Ferromagnetic \\
                    & \At phase & Anti-ferromagnetic \\
                    & B$_2$ phase & Homogeneous precession~\cite{bunkov2013spin} \\ 
        \hline
    \end{tabular}
    \label{tab:superfluid_phases}
\end{table}


\subsection{Phases of superfluid \he}\label{subsec:Hephasestructure}

Historically, after the success of the BCS theory \cite{PhysRev.108.1175}, people tried to look for the description of the superfluid \he because it is liquid and has no lattice structure inside. Some people considered the pairing states which are not $s$-wave. One is about the general anisotropic case by Anderson and Morel \cite{PhysRev.123.1911}. This model has a peculiar feature that the nodes exist on the Fermi surface for the axial $p$-wave state (refered to as the ABM state named after Anderson, Brinkman, and Morel). It turned out that this theory describes what is called the A phase nowadays. 
Later, Balian and Werthamer showed that the mixing of all substates of the $p$-wave Cooper pair is favored energetically \cite{PhysRev.131.1553}.
This state has an isotropic energy gap unlike the ABM state and is called the BW state, which is now recognized as the B phase. 
Experimentally, the A and B phases were discovered at $\SI{2.6}{mK}$ and $\SI{1.8}{mK}$ respectively \cite{PhysRevLett.29.920}, which confirmed the existence of the phase structure of the superfluid $^3$He.

The nucleus of a \he atom consists of two protons and one neutron.
The proton spins are aligned anti-parallel with each other, while the neutron spin is isolated, making the total spin angular momentum to be $1/2$.
In the superfluid phase, two \he atoms form a Cooper pair, whose ground state is a spin-triplet $p$-wave condensate~\cite{Vollhardt1990TheSP}.
The corresponding order parameter is expressed in terms of annihilation operators of nuclei $\hat{a}_{\vec{k}\alpha}$ as
\begin{equation}
    \expval{\hat{a}_{-\vec{k}\beta}\hat{a}_{\vec{k}\alpha}} \propto \Delta_{\vec{k} \alpha\beta} \equiv \sum_{\mu=1}^3 d_\mu (\vec{k}) (\sigma_\mu i \sigma_2)_{\alpha\beta}\, ,
\end{equation}
where $\vec{k}$ and $\alpha$ ($\beta$) are the momentum and the spin of a \he nucleus, respectively, and $\sigma_\mu$ is the Pauli matrix.
Since a Cooper pair forms a spin-triplet $L=1$ relative angular momentum state, the vector $d_\mu (\vec{k})$ can be represented as a linear combination of spherical harmonics $Y_{1m}(\vec{k}/|\vec{k}|)\propto \vec{k}/|\vec{k}|$,
\begin{equation}
    d_\mu(\vec{k}) = \sqrt{3} \sum_{j=1}^{3} A_{\mu j} \frac{\vec{k}_j}{|\vec{k}_j|} .
\end{equation}

The phenomenological Lagrangian of the Cooper pairs, i.e., Ginzburg--Landau Lagrangian, can be expressed in terms of the $3\times 3$ order parameter matrix $A_{\mu j}$ \cite{PhysRevA.8.2732,PhysRevLett.30.1135}. 
The index $\mu=1,2,3$ refers to the $S=1$ states while $j=1,2,3$ to the $L=1$ states, both in the Cartesian basis. Namely $A_{\mu j}$ transforms as a bi-vector under $\SO{3}_L \times \SO{3}_S$. Note that $A_{\mu j}$ is complex as its phase $\U{1}_\phi$ corresponds to the conserved number operator of the Cooper pairs.
Because the Lagrangian has to be Hermitian and invariant under the global $\SO{3}_L \times \SO{3}_S \times \U{1}_\phi$ symmetry,
we have only one second-order term of $A_{\mu j}$
\begin{equation}
    I_0 = \tr(AA^\dag) ,
    \label{eq:I_0}
\end{equation}
and five fourth-order terms
\begin{align}
    I_1 &= \abs{\tr(AA^T)}^2 ,
    \label{eq:I_1}\\
    I_2 &= \qty[\tr(AA^\dag)]^2 ,
    \label{eq:I_2}\\
    I_3 &= \tr[(AA^T)(AA^T)^*] ,
    \label{eq:I_3}\\
    I_4 &= \tr[(AA^\dag)^2] ,
    \label{eq:I_4}\\
    I_5 &= \tr[(AA^\dag)(AA^\dag)^*] ,
    \label{eq:I_5}
\end{align}
in the effective potential.
As a result, in the absence of any external fields, the effective potential per volume is given by
\begin{equation}
    V_0 = \alpha(T) I_0 + \frac{1}{2} \sum_{i=1}^{5} \beta_i I_i\, ,
    \label{eq:potential}
\end{equation}
where we neglect higher-order terms of $A_{\mu j}$, which can be justified when we consider the phenomenology of a system sufficiently close to the phase transition, and the numerical values of $|A_{\mu j}|$ are small.
The coefficients $\alpha$ and $\beta_i$ are determined by the microscopic theory.
For example, they have been calculated in the weak-coupling theory~\cite{Vollhardt1990TheSP}, and their numerical values are
\begin{align}
    & \alpha(T) \sim - 10^{-3} \qty(1-\frac{T}{T_c})\; \si{\micro eV^{-1} \angstrom^{-3}} \, ,\label{eq:alpha}\\
    & (\beta_1^{\mathrm{WC}}, \beta_2^{\mathrm{WC}}, \beta_3^{\mathrm{WC}}, \beta_4^{\mathrm{WC}}, \beta_5^{\mathrm{WC}}) = \frac{6}{5} \beta_0 \qty(-\frac{1}{2}, 1, 1, 1, -1) \, ,\label{eq:beta}\\
    & \beta_0 \sim 10^{-3} \; \si{\micro eV^{-3}\angstrom^{-3}} \, , \label{eq:beta0}
\end{align}
where $T_c$ is the transition temperature $\sim \SI{2.6}{mK}$ in the absence of external magnetic fields.
The values of $\beta_i$ can differ from those of $\beta^\mathrm{WC}_i$ depending on pressure.
Nevertheless, we will use the numerical values in \cref{eq:beta,eq:beta0} for $\beta_i$ below since the experimentally measured values differ from $\beta^\mathrm{WC}_i$ by only $\order{1}$ factors, $(\beta_i-\beta_i^{\mathrm{WC}})/\beta_0 = \mathcal{O}(1)$~\cite{choi2007strong}.

As noted above, the effective Lagrangian has a global symmetry $\SO{3}_L \times \SO{3}_S \times \U{1}_\phi$,  which corresponds to the rotation in the momentum space, the rotation in the spin space, and the overall phase rotation, respectively.
It is known that, depending on the values of coefficients in \cref{eq:potential}, the matrix $A$ acquires a non-zero expectation value in the ground state, which spontaneously breaks the global symmetry and leads to different phases.
Without an external magnetic field, there are two superfluid phases for \he, the A and B phases.
Their expectation values are expressed as
\begin{align}
    & \text{A phase: } A_{\mu j} \propto \frac{1}{\sqrt{2}} \mqty(0 & 0 & 0 \\ 0 & 0 & 0 \\ 1 & i & 0)\, , 
    \label{eq:A_phase} \\
    & \text{B phase: } A_{\mu j} \propto \frac{1}{\sqrt{3}} e^{i\phi} R_{\mu j}(\vec{n},\ \theta)\, ,
\end{align}
where $\phi$ is an overall phase, and $R_{\mu j}$ is a relative rotation of the spin and orbital spaces, represented by a rotation axis $\vec{n}$ and a rotation angle $\theta$.
Note that there are more than one choice of the order parameter in the A phase corresponding to the choices of particular directions of spin and orbital spaces, both of which are assumed to be the $z$-axis in the above expression.

When we turn on an external magnetic field $\vec{B}$, the potential $V$ has two more invariant terms
\begin{align}
    F^{(1)} &= i\eta \sum_{\mu\nu\lambda j}\epsilon_{\mu\nu\lambda} B_\mu A_{\nu j}^* A_{\lambda j} \, , \label{eq:F_1}\\
    F^{(2)} &\propto \sum_{\mu\nu j} B_\mu A_{\mu j} B_\nu A_{\nu j}^* \, .
\end{align}
Note that the magnetic field $\vec{B}$ couples with $A_{\mu j}$ only through the spin indices $\mu,\nu$ because the $^3$He atoms are electrically neutral, and their orbital angular momentum does not have a magnetic moment, while their spin angular momentum does.
Assuming that $\vec{B}$ is along the $z$-direction, one can see that $F^{(1)}$ and $F^{(2)}$ break the global symmetry to $\SO{3}_L \times \U{1}_{S_z}\times \U{1}_\phi$.
Because these interaction terms $F^{(1)}$ and $F^{(2)}$ bring three types of spontaneous symmetry breaking depending on the coefficients, there are three corresponding phases:
\begin{align}
    & \text{\Ao phase: } A_{\mu j} \propto \frac{1}{2} \mqty(1 & i & 0 \\ i & -1 & 0 \\ 0 & 0 & 0)\, , 
    \label{eq:A1_phase} \\
    & \text{\At phase: } A_{\mu j} \propto \frac{1}{\sqrt{2(\abs{p_1}^2 + \abs{p_2}^2)}} \mqty(p_1 & ip_1 & 0 \\ ip_2 & -p_2 & 0 \\ 0 & 0 & 0)\, , 
    \label{eq:A2_phase} \\
    & \text{\Bt phase: } A_{\mu j} \propto \frac{e^{i\phi}}{\sqrt{2\qty[2(\abs{p_1}^2+\abs{p_2}^2)+\abs{p_3}^2]}} \mqty(p_1 & p_2 & 0 \\ \pm p_2 & \mp p_1 & 0 \\ 0 & 0 & p_3)\, ,
\end{align}
where the real parameters $p_1$, $p_2$, and $p_3$ are uniquely determined as functions of the coefficients $\alpha(T)$ and $\beta_i$, as demonstrated in the next section.

\begin{figure}
    \centering
\includegraphics[width=12cm]{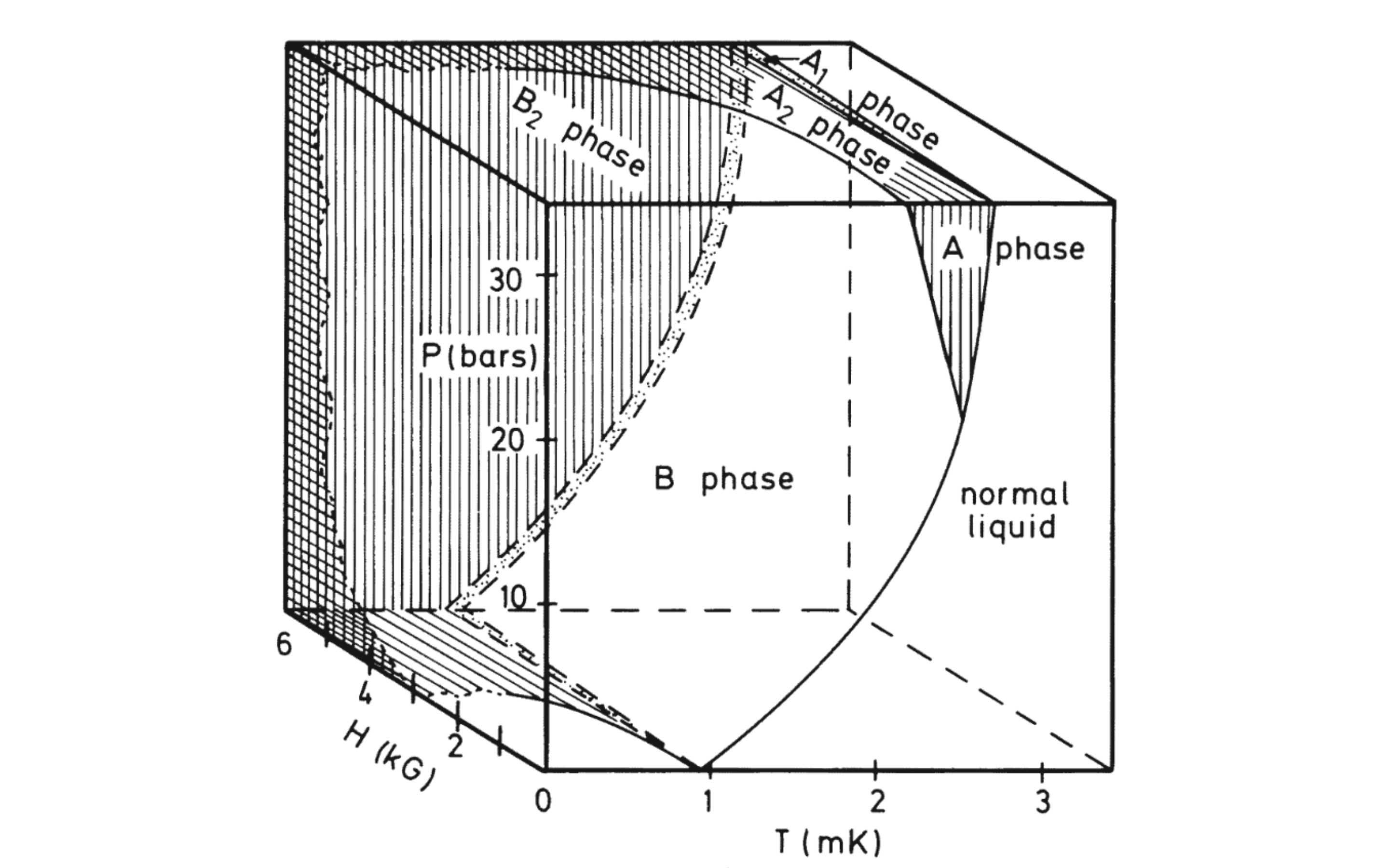}
    \caption{The $P$ (pressure) - $T$ (temperature) - $H$ (external magnetic field) phase diagram of \he taken from ref.~\cite{Vollhardt1990TheSP}.}
    \label{fig:phase3D}
\end{figure}

The $P$ (pressure) - $T$ (temperature) - $H$ (magnetic field) phase diagram of \he has been well studied~\cite{Vollhardt1990TheSP}, and we show it in \cref{fig:phase3D}.
It should be noted that the \Bt phase can only exist under relatively weak external magnetic fields, $H \lesssim \SI{0.5}{T}$, while the \Ao and \At phases can exist under strong magnetic fields, especially even for $H= \order{10}\;\si{T}$.
In the next section, we will see in more detail the criteria for which phase is realized, focusing on the A, \Ao, and \At phases.

\subsection{Spinor BEC description of magnetism in the A, \Ao, and \At phases}\label{subsec:HespinorBEC}

Hereafter, we focus on the A, \Ao, and \At phases, which have a unified description with the so-called spinor BEC formalism by keeping only the spin degrees of freedom. The spinor BEC refers to a Bose--Einstein condensate of atoms with integer spin, see e.g.\/, for a review~\cite{kawaguchi2012spinor,RevModPhys.85.1191}. 
This procedure is appropriate partly because the unbroken symmetries of these phases do not mix the rotations in spin and orbital spaces unlike the $B$ phases.
Thus, if we do not consider excitation of the orbital angular momentum of \he, we can focus only on the spin space.
For this purpose, we define a {\it spinor} order parameter $\vec{c}$ by fixing $L_z = +1$ as
\begin{equation}
    A_{\mu j} = \frac{1}{\sqrt{2}} \qty(\vec{c} ,\  i\vec{c} ,\ 0) \, \label{eq:spinor_c}.
\end{equation}
We can rewrite invariant terms $I_i$ and $F^{(1)}$ in terms of $\vec{c}$ as
\begin{align}
    I_0 &=
    \vec{c}^{\; *} \cdot \vec{c} \, , \\
    I_2 &=
    (\vec{c}^{\; *} \cdot \vec{c})^2 \, ,\\
    I_4 &=
    (\vec{c}^{\; *} \cdot \vec{c})^2 \, ,\\
    I_5 &=
    \abs*{\vec{c} \cdot \vec{c}}^2 = \qty[(\vec{c}^{\; *} \times \vec{c})^2 + (\vec{c}^{\; *} \cdot \vec{c})^2] \, , \\
    F^{(1)} & = i\eta \vec{B}\cdot(\vec{c}^{\; *} \times \vec{c}) \, .
    \label{eq:F1_phi}
\end{align}
Here, we do not consider the invariants $I_1$, $I_3$, and $F^{(2)}$ because these terms vanish for the \Ao and \At phases.
Finally, we get a simplified effective potential with the external magnetic field
\begin{align}
    V = \alpha (T) (\vec{c}^{\; *} \cdot \vec{c}) + \frac{\beta_{245}}{2} (\vec{c}^{\; *} \cdot \vec{c})^2 + \frac{\beta_5}{2} (\vec{c}^{\; *} \times \vec{c})^2+i\eta \vec{B}\cdot(\vec{c}^{\; *} \times \vec{c}) \, .
    \label{eq:GL_potential}
\end{align}
Here, we have defined a new parameter,
\begin{equation}
    \beta_{245} \equiv \beta_2 + \beta_4 + \beta_5\, .
\end{equation}
Note that $\beta_{245}>0$ and $\beta_5<0$ according to \cref{eq:beta}.
In the following, we discuss the magnetism of the \Ao and \At phases with this potential.

Using the simplified effective potential, we can easily analyze the potential form as a function of parameters.\footnote{
    Note that there can be a deeper minimum of the potential, which cannot be described by the spinor BEC formalism.
    Such a phase may correspond to the B or \Bt phase due to the spin-orbit couplings which originate from a long-distance dipole-dipole interaction among magnetic moments. However this effect is small and can be ignored in the presence of a strong magnetic field.
    It is worth noting, however, that any of the A, \Ao, and \At phases can be a global minimum of $V$ for reasonable choices of temperature, external magnetic field, and pressure, such as $T\simeq T_c$ and $B_z =\mathcal{O}(1)\,\si{T}$ under the standard atmosphere.
}
In the absence of an external magnetic field, only the temperature plays an important role.
For $T>T_c$, since $\alpha(T) > 0$ according to \cref{eq:alpha}, the potential $V$ has a global minimum at $\vec{c}=\vec{0}$, while for $T<T_c$ or $\alpha(T) < 0$, there is a global minimum at $\vec{c} \propto (0, 0, 1)^T$ with the potential energy $-\alpha^2/(2\beta_{245}) < 0$.
The former corresponds to the normal liquid phase, while the latter is consistent with the matrix structure of the A-phase order parameter \eqref{eq:A_phase}.

\begin{figure}
    \centering
    \includegraphics{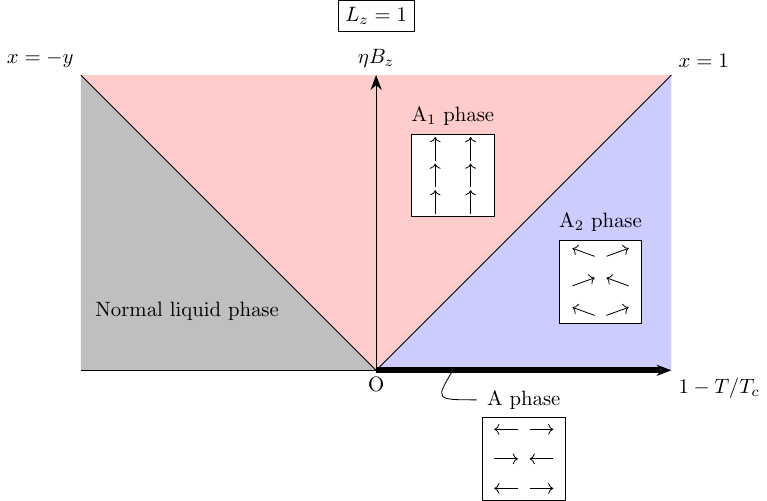}
    \caption{The schematics of the phase diagram focusing on A, \Ao, and \At phases of the superfluid \he. Here we fix the orbital angular momentum at $L_z=1$, so the B and \Bt phases do not appear in this phase diagram. The white box in each phase schematically represents the spin configuration of the Cooper pairs with the magnetic field $\vec{B}$ pointing {\it down} because of the negative $g$-factor.
    Note that the spins are not equally spaced as shown in this figure since the \he is not a solid in our setup.}
    \label{fig:phase}
\end{figure}

Next, we turn on the external magnetic field $\vec{B}=(0,0,B_z)^T$ with $\eta B_z>0$.
Restricting the form of $\vec{c}$ to be $(p_1,ip_2,0)^T$ with $p_1,p_2\in \mathbb{R}$, we obtain local minima of $V$ expressed as
\begin{align}
    &V=0 &\quad &\text{at} \quad \vec{c}=\vec{0}\, , \\
    &V=V_1 \equiv -\frac{\alpha(T)^2}{2 \beta_{245}}\frac{(x+y)^2}{y(1+y)} &\quad &\text{at} \quad \vec{c}=\vec{c}_1 \equiv \sqrt{\frac{-\alpha(T)(x+y)}{2\beta_{245}(1+y)}} \mqty(1 \\ i \\ 0)\, ,
    \label{eq:V1}\\
    &V=V_2 \equiv -\frac{\alpha(T)^2}{2 \beta_{245}} \frac{x^2+y}{y} &\quad &\text{at} \quad \vec{c}=\vec{c}_2 \equiv \sqrt{\frac{-\alpha(T)}{2\beta_{245}}} \mqty( \sqrt{1+\sqrt{1-x^2}} \\ i\sqrt{1-\sqrt{1-x^2}} \\ 0 )\, ,
    \label{eq:V2}
\end{align}
where we defined dimensionless variables
\begin{align}
    &x\equiv \frac{\beta_{245}\eta B_z}{\alpha(T) \beta_5} \propto B_z \qty(1-\frac{T}{T_c})^{-1}\, ,\\
    &y\equiv -\frac{\beta_{245}}{\beta_5}\, >0 \, .
\end{align}
The value of $x$ determines which of the local minima is the global minimum of $V$ as shown in \cref{fig:phase}.
Note that for $p_1\in \mathbb{R}$, the local minimum $\vec{c}=\vec{c}_1$ exists only when $x < -y$ or $x > 0$.
Similarly, the local minimum $\vec{c}=\vec{c}_2$ exists when $0 < x < 1$.
When $0<x<1$, we have $V_1 \geq V_2$, and this region corresponds to the \At phase (the blue region of \cref{fig:phase}).
When $x > 1$ or $x < -y$, we have $V_2 \geq V_1$, which corresponds to the \Ao phase (the red region).
When $-y<x<0$, we obtain the normal liquid phase (the gray region).

For later convenience, we define a normalized order parameter
\begin{equation}
    \vec{\phi} \equiv \frac{\sqrt{\ncp}}{\Delta} \vec{c}\, ,
    \label{eq:phi_def}
\end{equation}
where $\Delta$ is a normalization factor with a dimension of energy defined as
\begin{align}
    \Delta \equiv \sqrt{\vec{c}^{\; *} \cdot \vec{c}} \, ,
\end{align}
so that $\vec{\phi}^*\cdot\vec{\phi} = \ncp$ with $\ncp$ being the number density of Cooper pairs.
Using the typical size of the gap energy $E_g\sim \SI{e-6}{eV}$ and the Fermi energy $E_F\simeq \SI{0.9e-4}{eV}$, the number density of Cooper pairs $\ncp$ can be estimated as
\begin{align}
    \ncp \sim \frac{E_g}{E_F} \frac{n_{^3\mathrm{He}}}{2} \sim 10^{-2} \frac{n_{^3\mathrm{He}}}{2},
    \label{eq:Cooper_density}
\end{align}
where $n_{^3\mathrm{He}} \simeq \SI{2e-2}{\angstrom^{-3}}$ is the number density of \he atoms \cite{godfrin2022dynamics,PhysRevLett.46.728,PhysRevLett.62.1130,RevModPhys.51.821,PhysRevB.28.3770}. 
\Cref{eq:Cooper_density} is consistent with the experimental value \cite{kojima2008spin} that shows superfluid density fraction is $\order{10^{-2}}$. 
The effective potential is now written in terms of $\vec{\phi}$ as
\begin{align}
    V=-\mu\vec{\phi}^*\cdot\vec{\phi}
    +\frac{\mu}{2v^2}(\vec{\phi}^*\cdot\vec{\phi})^2
    -\lambda(\vec{\phi}^*\times\vec{\phi})^2+i g\mu_N \vec{B}\cdot (\vec{\phi}^*\times\vec{\phi}).
    \label{eq:V}
\end{align}
Here we have defined some new parameters $\mu \equiv -\alpha(T) \Delta^2/n_\mathrm{C}$, $v^2 \equiv - \alpha(T) n_\mathrm{C}/ (\beta_{245} \Delta^2)$, and $\lambda \equiv - \beta_5 \Delta^4 / (2n_\mathrm{C}^2)$.
Typical sizes of parameters are estimated as $\mu\sim \si{neV}$, $v\sim \si{\angstrom^{-3/2}}$, and $\lambda\sim \si{neV \angstrom^3}$.
In the last term of the potential, $g \simeq -4.3$ is the $g$-factor of the \he nucleus~\cite{Schneider:2022mze}, while $\mu_N \simeq 3.2 \times 10^{-8}\,\mathrm{eV}\,\mathrm{T}^{-1}$ is the nuclear magneton.
This choice of the coefficient is justified by the fact that the spin density is expressed as $\vec{s} \equiv -i (\vec{\phi}^* \times \vec{\phi})$.
Indeed, the last term describes the interaction between the magnetic field and the spin of the form $g \mu_N \vec{B}\cdot \vec{s}$.

We can now study the ordering of nuclear spins using $\vec{\phi}$ and its expectation values in different phases.
In the \At phase, the spin per Cooper pair is calculated as
\begin{equation}
    \vec{S} \equiv \frac{\vec{s}}{\ncp} = \mqty(0 \\ 0 \\ x)\, .
\end{equation}
In the limit of $B_z \to 0$ or $x\to 0$, this phase is smoothly connected to the A phase, which has an anti-ferromagnetic ordering with $\vec{S}=\vec{0}$.
In the \Ao phase, the spin per Cooper pair is
\begin{equation}
    \vec{S} = \mqty(0 \\ 0 \\ 1)\, ,
\end{equation}
which shows that the spins of Cooper pairs are completely aligned along the direction of $g\mu_N \vec{B}$.
Therefore, we conclude that the \Ao phase has a ferromagnetic ordering.

\subsection{Nuclear magnons in the ferromagnetic \Ao phase}\label{subsec:nuclearmagnon}

Depending on the symmetry-breaking patterns in different phases, there appear several gapless modes, the so-called Nambu-Goldstone (NG) modes.
These modes are classified as type-A and type-B modes with characteristic dispersion relations at the long-wavelength limit  \cite{Watanabe:2012hr,Hidaka:2012ym}.
For example, in the ferromagnetic \Ao phase, the coset space is given by
\begin{align}
\mathbb{R}P^3=\frac{\mathrm{SO}(3)_S\times\mathrm{U}(1)_\phi}{\mathrm{SO}(2)_{S_z-\phi}},
\end{align}
which corresponds to one type-A NG mode with a linear dispersion and one type-B NG mode with a quadratic dispersion.
The type-B mode is identified as an acoustic magnon mode, whose gap can be generated by the soft symmetry-breaking effect, including the external magnetic field.
On the other hand, in the anti-ferromagnetic \At phase, the coset space is given by
\begin{align}
S^2\times\mathrm{U}(1)_\phi=\frac{\mathrm{SO}(3)_S\times\mathrm{U}(1)_\phi}{\mathrm{SO}(2)_{S_z}},
\end{align}
which corresponds to three type-A NG modes, two of which are identified as magnon modes with 
$S_x$ and $S_y$.
Since the magnon modes in the ferromagnetically-ordered phase have the strongest interaction with the spatially uniform magnetic field, such as the one induced by the axion dark matter, we will focus on the type-B magnon mode in the \Ao phase.

The excitation modes in the superfluid \he can be studied by treating the normalized order parameter $\vec{\hat{\phi}}$ as a dynamical field.
The field theory Lagrangian is given by
\begin{align}
    \mathcal{L}=i\vec{\hat{\phi}}^\dag \cdot\partial_t\vec{\hat{\phi}}
    -\frac{1}{2m^\star}\sum_i (\partial_i\vec{\hat{\phi}}^\dag)\cdot(\partial_i \vec{\hat{\phi}})
    -V,
    \label{eq:L_phi}
\end{align}
where $i=x,y,z$ are the space coordinates, and the potential $V$ is given by \cref{eq:V}.
The effective mass $m^\star$ depends on the pressure imposed on \he and can be experimentally determined through measurements of the specific heat.
The typical value of $m^\star$ is about 3 to 6 times larger than the $^3$He atomic mass \cite{PhysRevB.33.7520}.

In order to study the magnon excitation mode in the \Ao phase,
we add a quantum fluctuation $\hat{\psi}$ to the expectation value $\langle\vec{\hat{\phi}}\rangle = \sqrt{\ncp/2}(1, i, 0)$ as
\begin{equation}
   \vec{\hat{\phi}} = \frac{\sqrt{\ncp}}{2\sqrt{2}} \mqty(
       2-\hat{\psi}^\dagger \hat{\psi} - \hat{\psi}^2 \\
       i (2-\hat{\psi}^\dagger \hat{\psi} + \hat{\psi}^2) \\
       - 2 \sqrt{2-\hat{\psi}^\dagger \hat{\psi}} \, \hat{\psi}
   ).
\end{equation}
We also consider the fluctuation of the magnetic field as $\vec{B} = (0,0,-B_z)^T + \delta \vec{B}$ with $B_z>0$.
For simplicity, we assume that both $\hat{\psi}$ and $\delta\vec{B}$ do not depend on the space coordinate.
By substituting the expansion in the Lagrangian \eqref{eq:L_phi} and picking up only the leading-order terms of the fluctuation $\hat{\psi}$ and $\delta\vec{B}$, we obtain the following terms
\begin{align}
    \delta\mathcal{L}
    &= g\mu_N \ncp B_z \hat{\psi}^\dag \hat{\psi} + \frac{1}{\sqrt{2}} g \mu_N \ncp
    \qty(\delta B_x(\hat{\psi}+\hat{\psi}^\dagger)- i \delta B_y(\hat{\psi}-\hat{\psi}^\dagger)),
    \label{eq:L_psi}
\end{align}
which originally come from the last term of the potential \eqref{eq:V}.

It is convenient to discuss in terms of the non-relativistic Hamiltonian described with the magnon operators.
For this purpose, we first obtain the relationship of the spin density
\begin{align}
    \hat{s}_+ &\equiv \hat{s}_x + i \hat{s}_y = \ncp \sqrt{2-\hat{\psi}^\dag\hat{\psi}} \; \hat{\psi}, \\
    \hat{s}_- &\equiv \hat{s}_x - i \hat{s}_y = \ncp \hat{\psi}^\dagger \sqrt{2-\hat{\psi}^\dag\hat{\psi}}, \\
    \hat{s}_z &= \ncp (1-\hat{\psi}^\dag\hat{\psi}).
\end{align} 
On the other hand, using the Holstein--Primakoff transformation with the spin size $s=1$, we can relate the spin operator of each Cooper pair labeled by $\ell$ to the magnon annihilation and creation operators as
\begin{align}
    \hat{S}_\ell^{+} &= \sqrt{2-\hat{b}_\ell^\dagger \hat{b}_\ell}\, \hat{b}_\ell , \\
    \hat{S}_\ell^{-} &= \hat{b}_\ell^\dagger \sqrt{2- \hat{b}_\ell^\dagger \hat{b}_\ell} , \\
    \hat{S}_\ell^z &= 1 - \hat{b}_\ell^\dagger \hat{b}_\ell ,
\end{align}
with the canonical commutation relation of bosonic operators $[\hat{b}_\ell, \hat{b}_{\ell'}^\dagger] = \delta_{\ell \ell'}$.
We are only interested in the spatially uniform mode obtained by the Fourier transformation $\maga \equiv \sum_{\ell=1}^{\Ncp} \hat{b}_\ell / \sqrt{\Ncp}$, where $\Ncp \equiv \ncp V_{^3\mathrm{He}}$ is the total number of Cooper pairs with $V_{^3\mathrm{He}}$ being the volume of the superfluid \he.
We find that this mode is related to the spatially uniform fluctuation $\hat{\psi}$ as
\begin{align}
    \maga = \sqrt{\Ncp} \hat{\psi}.
    \label{eq:magnon_spinor_relation}
\end{align}
Note that \cref{eq:magnon_spinor_relation} is consistent when $\hat{\psi}$ obeys a bosonic commutation relation, which is the case for the spinor BEC formalism.

Finally, substituting the magnon operator \eqref{eq:magnon_spinor_relation} in the Lagrangian \eqref{eq:L_psi}, we obtain the relevant part of the Hamiltonian
\begin{align}
    H = \omega_L \magc \maga - 
    \sqrt{\frac{\Ncp}{2}} g \mu_N \left(
        \delta B_x (\maga + \magc)
        - i \delta B_y (\maga - \magc)
    \right) + \cdots,
    \label{eq:Bint_magnon}
\end{align}
where $\omega_L \equiv -g \mu_N B_z$ is the Larmor frequency.
As we will see below, the second term causes the magnon excitation by the axion-induced effective magnetic field.

\section{Axion detection}\label{sec:Axiondetection}

In this section, we explain the details of our axion detection experiment using superfluid \he.
A brief overview is as follows: (i) axions excite the magnon modes in the \Ao phase of \he; (ii) these magnons mix with cavity photon modes; (iii) the signal photons are amplified and detected.
We also discuss the amplification of the signal using quantum measurement techniques.

\subsection{Axion-magnon conversion}\label{subsec:Axionnuclear}

As is mentioned above, the spin angular momentum of a \he nucleus originates from the neutron spin.
As a result, the axion-proton coupling can be neglected in our discussion, which generally has a different value from the axion-neutron coupling.
The axion-neutron dynamics is described by the Lagrangian
\begin{align}
  \mathcal{L} = \frac{1}{2} (\partial_\mu a)^2 - \frac{1}{2} m_a a^2
  + \bar{n} (i\Slash{\partial} - m_n) n
  + C_{ann} \frac{\partial_\mu a}{2f_a} \bar{n}\gamma^\mu\gamma_5 n,
  \label{eq:magnon}
\end{align}
where $a$ and $n$ are the axion and the neutron fields with masses $m_a$ and $m_n$, respectively, $C_{ann}$ is a model-dependent $\order{1}$ coupling coefficient, and $f_a$ is the axion decay constant.
For the QCD axion, there is a relationship between $m_a$ and $f_a$~\cite{Gorghetto:2018ocs}:
\begin{align}
    m_a \simeq \SI{5.7}{\micro eV} \left(
        \frac{\SI{e12}{GeV}}{f_a}
    \right).
\end{align}

We assume that the axion field explains all of the dark matter abundance through the misalignment mechanism~\cite{Preskill:1982cy,Abbott:1982af,Dine:1982ah}; accordingly, the axion field can be treated as a classical field with coherent oscillation
\begin{align}
  a(t, \vec{x}) \simeq a_0 \sin(m_a t - m_a \vec{v}_a \cdot \vec{x} + \varphi),
\end{align}
where $v_a$ is the velocity of axion, while $\varphi$ is a random phase.
Here, we utilize the fact that the axion is non-relativistic to approximate the axion energy to be $m_a$.
Using these variables, the local dark matter density $\rho_a\sim \SI{0.45}{GeV / cm^3}$ can be expressed as $\rho_a=(m_a a_0)^2/2$.
The expression of $a(t,\vec{x})$ tells us that the coherent length of the axion field is given by $\lambda_a \equiv 1/(m_a v_a)$.
Since $\lambda_a \sim 100\,\mathrm{m}$ for $m_a = \SI{1}{\micro eV}$ and $v_a \sim 10^{-3}$~\cite{particle2022review}, the axion field can be regarded as a spatially uniform field within an experimental apparatus, which allows us to neglect the second argument of the sine function.
Also, the coherence time of the axion field is $\tau_a \simeq 1/(m_a v_a^2) \sim \SI{1}{ms}$ for $m_a = \SI{1}{\micro eV}$, during which the velocity $\vec{v}_a$ and the phase $\varphi$ can be treated as constant.

In the non-relativistic limit, we obtain the following effective Hamiltonian density describing the axion-nucleus interaction:
\begin{align}
     \mathcal{H}_{\mathrm{eff}}\simeq - C_{ann} \frac{m_a a_0}{f_a} \vec{v}_a \cdot \vec{s}_N \sin(m_a t + \varphi),
\end{align}
where $\vec{s}_N$ is the spin density operator of $^3$He nuclei, which can be identified as the spin operator of neutrons in the \he.
Note that the interaction strength is proportional to $m_a a_0 = \sqrt{2 \rho_a}$ and independent of $m_a$.
The interaction term can be rewritten in the form of the ordinary spin-magnetic field coupling, $\mathcal{H} = \gamma_N \vec{B}_a \cdot \vec{s}_N \sin(m_a t + \varphi)$, where $\gamma_N = g \mu_N$ is the gyromagnetic ratio of a nucleus.
The effective axion magnetic field that exclusively couples to the neutron spins is given by
\begin{align}
  \gamma_N \vec{B}_a(t) = - C_{ann} \frac{\sqrt{2 \rho_a}}{f_a} \vec{v}_a \sin(m_a t + \varphi).
\end{align}
Thus, by substituting $\delta \vec{B}$ by $\vec{B}_a$ in \cref{eq:Bint_magnon}, we obtain the Hamiltonian of the axion-nuclear magnon coupled system
\begin{align}
    H(t) &= H_0 + H_{\mathrm{int}}(t), \\
    H_0 &= \omega_L \magc \maga, \\
    H_{\mathrm{int}}(t) &= \frac{C_{ann}}{f_a} \sqrt{\rho_a \Ncp} \left(
        v_a^{+} \magc + \mathrm{h.c.}
    \right) \sin(m_a t + \varphi),
    \label{eq:axion-magnon-int}
\end{align}
where $v_a^{+} \equiv v_a^x + i v_a^y$.

We define the ground state $\ket{0}$ and the one-magnon state $\ket{1}$ of \he with $\maga \ket{0} = 0$ and $\ket{1} \equiv \magc \ket{0}$.\footnote{
    States with more than one magnon can be safely neglected due to the smallness of the magnon excitation rate for the axion parameter region of our interest.
}
Then, the magnon production amplitude is calculated as
\begin{align}
    -i\mathcal{M}=\bra{1}U(t)\ket{0}
    =-i\int_0^t \mathrm{d}t'\, \Braket{1 | H_\mathrm{int}(t') | 0}
    e^{-i\omega_L t'},
\end{align}
where $t < \tau_a$ is the observation time and the evolution matrix is defined as
\begin{align}
    U(t) \equiv \exp \left[
        - i \int_0^t \mathrm{d}t'\, H(t')
    \right].
\end{align}
Since the axion spectrum is approximately monochromatic with energy $m_a$, the magnon production rate is resonantly enhanced when $m_a = \omega_L$.
In this limit, the amplitude is evaluated as
\begin{align}
    \mathcal{M}
    &\simeq -i\frac{C_{ann}}{2f_a} \sqrt{\rho_a \Ncp} v_a^{+} e^{i\varphi} t,
\end{align}
where we assumed $t\gg \omega_L^{-1}$ so that the oscillatory term can be dropped. Then the transition probability is
\begin{align}
    P=|\mathcal{M}|^2=\left(\frac{C_{ann}}{2f_a}\right)^2\rho_a \Ncp t^2 v_a^2 \sin^2\theta_a,
\end{align}
where $\theta_a$ is the relative angle between the external magnetic field and axion wind.
This result is consistent with ref.~\cite{Chigusa:2020gfs} where the spatially uniform mode (the Kittel mode) of the electronic magnons is considered.

The transition probability grows as $P\propto t^2$ as far as the coherence of the signal is maintained.
The typical coherence time $\tau$ can be estimated as
\begin{align}
    \tau \sim \min \left(
        \tau_a, \taum, \tau_{\mathrm{exp}}
    \right),
    \label{eq:typical_time}
\end{align}
where $\taum$ is the lifetime of magnons, and $\tau_{\mathrm{exp}}$ denotes the minimum relaxation time scale of excitation modes used for the magnon detection.
Since we use the mixing between a nuclear magnon and a cavity photon as is discussed in \cref{subsec:mixingmagcav}, the cavity quality factor $Q$ contributes to $\tau_{\mathrm{exp}}$ in the form of $Q/m_a$.
The magnon lifetime $\taum$ is identified as the spin relaxation time.
In general, there are two types of spin relaxation times; the longitudinal and transverse spin relaxation time $T_1$ and $T_2$, which characterize the relaxation of the longitudinal and transverse component of the magnetization vector, respectively.
For experiments that utilize nuclear magnetic resonance, such as our experiment, the crucial factor is $T_2$ as can be seen in \cref{eq:Bint_magnon}.
For the \Ao phase of superfluid \he, $T_1$ has been decided experimentally as an order of $\order{1\text{-}10}\;\si{s}$ \cite{corruccini1978pulsed,lu1989spin}.
However, $T_2$ has not been measured because of some experimental difficulties \cite{Murakawa2023}.
One of the difficulties in measuring the intrinsic $T_2$ is due to the inhomogeneity of the magnetic field, which is significant under a high magnetic field such as in the \Ao phase.
Another difficulty is caused by the effect called ``motional narrowing", which appears in an inhomogeneous medium such as liquid samples and makes $T_2$ longer.
Since $T_2$ is typically shorter than $T_1$ by an order of $\order{10^{-2}\text{-}10^{-1}}$, we use $\taum=\SI{1}{s}$ in this paper, which is much longer than the axion coherence time $\tau_a=\order{1}\;\si{ms}$ for the region of our interest.
We also assume $\tau_{a} < \tau_{\mathrm{exp}}$ and use $\tau = \tau_a$ for the following calculation, which is reasonable for $Q\gtrsim 10^6$.
Of course, before actually performing our experiment, the transverse spin relaxation time $T_2$ should be measured first in the fixed setup.

Finally, the signal rate for the total observation time $t \gg \tau$ is evaluated as
\begin{align}
    \dv{N_{\text{sig}}}{t}=\frac{\Ncp}{4}C_{ann}^2\frac{\rho_a v_a^2 \sin^2\theta_a}{f_a^2} \tau,
    \label{eq:signal_rate}
\end{align}
where $\sin^2\theta_a$ should be replaced by the averaged value if $t \gg \tau_a$.
Hereafter, we assume this is the case and simply average out the directional dependence, though it might be interesting to study it further in light of the modulation of the axion signal.
Note that the total number of Cooper pairs for superfluid \he of mass $M$ is calculated as $\Ncp\sim 10^{-2} M/(2m_{^3\mathrm{He}})\sim 1.0\times 10^{23} (M/\SI{100}{g})$ according to \cref{eq:Cooper_density}.
For the QCD axion, for example, the external magnetic field $B_z=\SI{10}{T}$ corresponds to the Larmor frequency $\omega_L = m_a\simeq \SI{1.3}{\micro eV}$ and $f_a\simeq \SI{4.3e12}{GeV}$, which result in
\begin{align}
    \dv{N_{\mathrm{sig}}}{t}=\SI{1.1e-5}{s^{-1}} \times
    C_{ann}^2 \left(
        \frac{M}{\SI{100}{g}}
    \right) \left(
        \frac{v_a}{10^{-3}}
    \right)^2 \left(
        \frac{\tau}{\SI{1}{ms}}
    \right) \, \sin^2\theta_a.
\end{align}
We also show the expression of the signal power:
\begin{align}
    P_\mathrm{sig} = \SI{2.2e-30}{W} \times C_{ann}^2
    \left(\frac{M}{\SI{100}{g}}\right)
    \left(\frac{v_a}{10^{-3}}\right)^2
    \left(
        \frac{\tau}{\SI{1}{ms}}
    \right) \, \sin^2\theta_a.
\end{align}

\subsection{Mixing between magnon and cavity modes}\label{subsec:mixingmagcav}

When one of the cavity modes has the same frequency as the magnons of our interest, $\omega_{\mathrm{cavity}} = \omega_L$, there is a large mixing between these modes.
This can be understood similarly as the formation of the magnon-polariton of electron spins \cite{PhysRevLett.113.156401, tabuchi2014hybridizing, tabuchi2016quantum}.
Let $\hat{c}$ ($\hat{c}^\dagger$) be the annihilation (creation) operator of the cavity mode.
Assuming that all the other cavity modes have frequencies largely deviated from $\omega_L$, we can safely neglect them and write down the relevant part of the Hamiltonian
\begin{align}
    H = \omega_L \magc \maga
    + \omega_{\mathrm{cavity}} \hat{c}^\dagger \hat{c}
    + H_{\mathrm{mix}}.
\end{align}
The mixing term is sourced from the interaction between nucleon spin and the magnetic field of the cavity mode and is given by
\begin{align}
    H_{\mathrm{mix}} = i g \mu_N \int_{^3\mathrm{He}} \mathrm{d}V
    \left(
        \vec{\phi}^* (\vec{r}) \times \vec{\phi} (\vec{r})
    \right)
    \cdot \vec{B}_0 (\vec{r}) (\hat{c} + \hat{c}^\dagger),
\end{align}
where the volume integral is performed over the volume of the superfluid $^3$He, while $\vec{B}_0 (\vec{r})$ is the profile of the magnetic field of the cavity mode.
If we consider as an example the cavity mode with $\vec{B}_0 (\vec{r}) = B_0(\vec{r}) \vec{u}_x$ with $\vec{u}_x$ being the unit vector along the $x$-axis, terms linear in the magnon mode is obtained similarly to \cref{eq:Bint_magnon} as
\begin{align}
    H_{\mathrm{mix}} \simeq \sqrt{\frac{\Ncp}{2}} g \mu_N \overline{B}_0
    (\maga + \magc) (\hat{c} + \hat{c}^\dagger) ,
\end{align}
where the averaged magnetic field over the superfluid \he is defined as
\begin{align}
    \overline{B}_0 \equiv \frac{1}{V_{^3\mathrm{He}}}
    \int_{^3\mathrm{He}} \mathrm{d}V\, B_0 (\vec{r}).
\end{align}
We finally find the quadratic part of the Hamiltonian
\begin{align}
    H &\simeq \omega_L \magc \maga + \omega_{\mathrm{cavity}} \hat{c}^\dagger \hat{c}
    + g_{\mathrm{eff}} (\hat{c} \magc + \hat{c}^\dagger \maga),
    \label{eq:polariton} \\
    g_{\mathrm{eff}} &= \sqrt{\frac{\Ncp}{2}}\, g \mu_N \overline{B}_0,
\end{align}
where we used the rotating wave approximation to neglect the fast oscillation terms.
Note that the typical size of the magnetic field can be estimated by matching the electromagnetic energy with a cavity mode frequency.
Defining $\left< B_0^2 \right> \equiv \frac{1}{V_{\rm cavity}} \int_{\mathrm{cavity}} \mathrm{d}V\, B_0^2(\vec{r})$ with integration over the cavity volume, we obtain
\begin{align}
    \sqrt{ \left< B_0^2 \right> } \sim 4 \,\mathrm{f T} \left(\frac{\omega_{\mathrm{cavity}}}{\SI{e2}{M Hz} }\right)^{1/2} \left(\frac{\SI{e3}{cm^{3}} }{V_{\mathrm{cavity}}}\right)^{1/2} .
\end{align}
For the order estimation of the physics scales, we can approximate that $\overline{B}_0 \sim \sqrt{\Braket{B_0^2}}$, though there can be an $\order{1}$ geometry factor difference.
Indeed, this estimation is consistent with ref.~\cite{tabuchi2016quantum}, which shows that $\overline{B}_0 \sim 5\,\mathrm{pT}$ in one of the figures, while a rough estimation gives $\sqrt{\Braket{B_0^2}} \sim 1\,\mathrm{pT}$.

By diagonalizing the Hamiltonian \eqref{eq:polariton}, we obtain the energy eigenstates.
In particular, the maximal mixing is realized when $\omega_L = \omega_{\mathrm{cavity}}$ with the corresponding energy eigenvalues $|\omega_L \pm g_{\mathrm{eff}}|$.
Compared with the magnon-polariton of electron spins, the energy scale of the system is smaller by a factor of $\mu_N / \mu_B \sim 10^{-3}$ with $\mu_B$ being the Bohr magneton.
This affects the time scale of the conversion of the magnon mode into the cavity mode.
The time scale can be estimated by evaluating the energy gap $\Delta E = 2\min \left( \omega_L=\omega_{\mathrm{cavity}}, g_{\mathrm{eff}} \right)$ between two energy eigenstates.
Assuming $V_{^3\mathrm{He}} \sim V_{\mathrm{cavity}}$ for simplicity, and the above estimation of $\overline{B}_0$, we have 
\begin{align}
    \frac{g_{\mathrm{eff}}}{2\pi} \sim 0.3\,\mathrm{MHz}\, 
    \left(
        \frac{M}{\SI{100}{g}}
    \right)^{1/2} \left(
        \frac{\omega_L}{10^{2} \,\mathrm{M Hz}}
    \right)^{1/2},
    \label{eq:g_eff}
\end{align}
where the $\omega_L = \omega_{\mathrm{cavity}}$ dependence comes from that of $\sqrt{\expval{B_0^2}}$.
This expression, together with $\omega_L \sim 200\,\mathrm{MHz}$ for $B=1\,\mathrm{T}$, shows that the conversion time scale, which is usually set by $g_{\mathrm{eff}}^{-1} \sim \si{\micro s}$, can be much shorter than the typical coherence time $\tau \sim \si{ms}$.
Thus, it is expected that half of magnons excited by the axion DM are converted to cavity modes, which can be observed by the following detector.
Note, however, that $\geff$ highly depends on the detector setup including its geometry, and should carefully be estimated once the setup is fixed.
In \cref{sec:SNR}, we show the detailed calculation of the dynamics of the magnon-cavity mixed system including various loss factors and quantum measurement techniques briefly introduced in the next subsection.

\subsection{Quantum measurement techniques}
\label{sec:QMT}

In the following, we consider two noise sources for our experimental setup according to the discussion in refs.~\cite{Malnou:2018dxn,HAYSTAC:2020kwv}.
The first is thermal noise, or Johnson--Nyquist noise, sourced from the internal loss of the cavity, and the second is thermal noise sourced from a termination resistor.
Each of their spectral densities is given by the formula
\begin{align}
    n_T + \frac{1}{2} \equiv \frac{1}{\exp(\omega/k_B T) - 1} + \frac{1}{2},
    \label{eq:thermal_photon}
\end{align}
where $T$ is the temperature of the cavity.
Even at zero temperature, the noise has a nonzero value known as quantum noise, which originates from the quantum fluctuation.
This is known as the standard quantum limit (SQL), and the noise floor is expressed in terms of a temperature~\cite{van2020putting}
\begin{equation}
    T_{\mathrm{SQL}} \equiv \frac{\omega}{k_B} \simeq \SI{12}{mK}\, \qty(\frac{\omega}{\SI{1}{\micro eV}}).
\end{equation}
Thus, the quantum noise dominates the thermal noise \eqref{eq:thermal_photon} for setups below $\SI{12}{mK}$ when we want to look for signals from $\SI{1}{\micro eV}$ axions.
For our setup, in which the cavity is cooled to about $\SI{2.6}{mK}$, the quantum noise dominates for the axion mass $m_a\gtrsim \SI{0.22}{\micro eV}$.
This quantum noise does not seem to be able to be further reduced under temperatures below $T_\mathrm{SQL}$, but this SQL can be circumvented by using quantum measurement techniques (see ref.~\cite{RevModPhys.82.1155} for a review).
Specifically, we use two quantum measurement techniques; the squeezing of states and the homodyne measurement, as introduced in ref.~\cite{Malnou:2018dxn}.
We will summarize these techniques in this section.

\subsubsection{Squeezing of states}
\label{sec:squeezing}
The starting point is introducing {\it quadratures} $\hat{X}$ and $\hat{Y}$ defined in terms of the annihilation (creation) operator of photons $\hat{a}$ ($\hat{a}^\dag$) as
\begin{equation}
    \hat{X} \equiv \frac{\hat{a} + \hat{a}^\dag}{\sqrt{2}}\, ,
    \quad
    \hat{Y} \equiv \frac{\hat{a} - \hat{a}^\dag}{\sqrt{2}i}\, .
    \label{eq:quadrature}
\end{equation}
Because of the commutation relation $[\hat{a}, \hat{a}^\dag] = 1$, quadratures satisfy
\(
    [\hat{X}, \hat{Y}] = i\, .
\)
This commutation relation results in the uncertainty relation of quadratures
\begin{equation}
    (\Delta \hat{X})^2(\Delta \hat{Y})^2 \ge \frac{1}{4}\, .
    \label{eq:uncertainty}
\end{equation}
Since many of the ordinary measurement techniques measure both quadratures of the input signal at each time, the quantum noise $\Delta\hat{X}\sim \Delta\hat{Y} \sim 1/2$ must appear and contribute to the SQL.
However, quantum measurement techniques can decrease this quantum noise by focusing on only one of the quadratures.
For example, a larger part of the uncertainty can be imposed on $\hat{Y}$, as $\Delta\hat{X}\sim 1/(2\sqrt{G})$ and $\Delta\hat{Y}\sim \sqrt{G}/2$ with $G\gg 1$, which reduces the uncertainty on the observable $\hat{X}$ and remains consistent with \cref{eq:uncertainty}.
This operation is called squeezing.
Squeezing can be performed by, e.g., phase-sensitive amplifiers such as Josephson parametric amplifiers (JPAs); see \cref{sec:JPA} for details.

\begin{figure}
    \centering
    \includegraphics[scale=0.85]{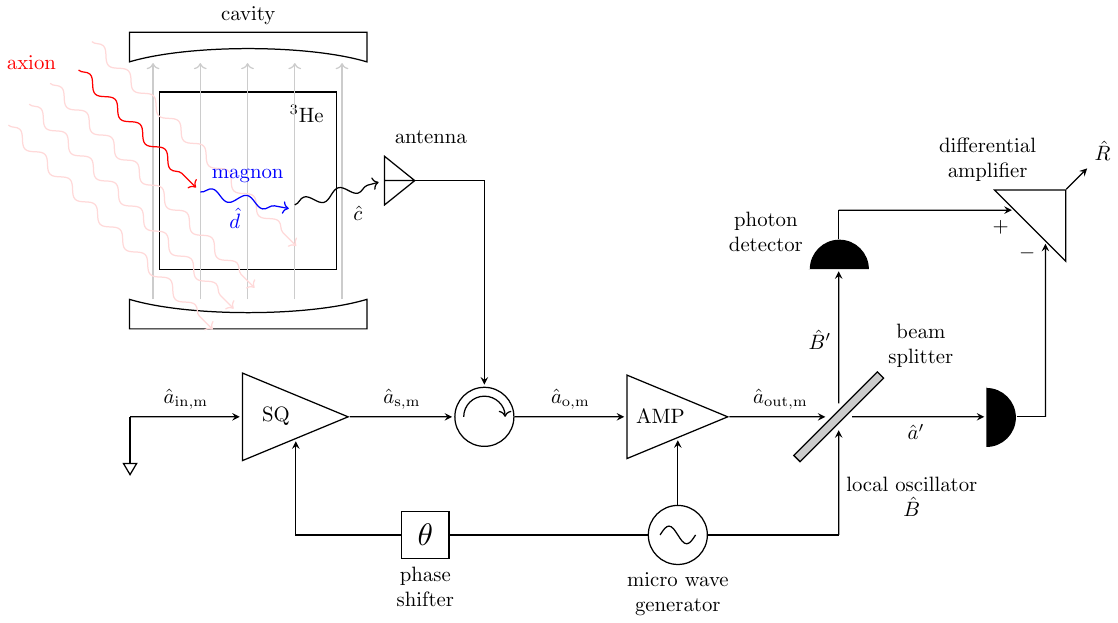}
    \caption{Schematic of our experimental setup for axion detection with superfluid \he. The operators $\hat{c}, \hat{d}, \hat{B}, \cdots$ correspond to the annihilation operators used in our paper.}
    \label{fig:setup}
\end{figure}

\begin{figure}
    \centering
    \includegraphics[scale=1]{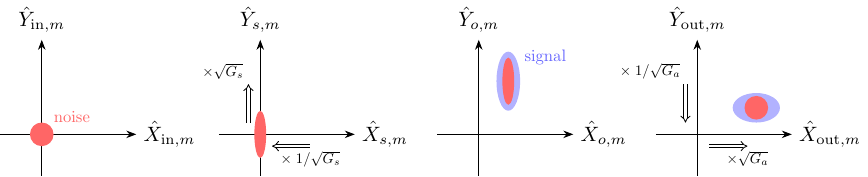}
    \caption{Distribution of the four states in the $XY$ plane. The subscripts of quadratures correspond to those in \cref{fig:setup}. The input state $(\hat{X}_{\text{in}, m}, \hat{Y}_{\text{in}, m})$ is Gaussian, which is the distribution of thermal photons. This state will be squeezed by the SQ JPA and becomes the squeezed state $(\hat{X}_{s, m}, \hat{Y}_{s, m})$. The third state $(\hat{X}_{o, m}, \hat{Y}_{o, m})$ is the state after the signal from the cavity is received. Finally, we get the output state $(\hat{X}_{\text{out}, m}, \hat{Y}_{\text{out}, m})$ after squeezing by the AMP JPA.}
    \label{fig:XYdist}
\end{figure}

A possible experimental setup, which is similar to the setup of the HAYSTAC experiment~\cite{HAYSTAC:2018rwy, HAYSTAC:2020kwv, HAYSTAC:2023cam}, is schematically shown in \cref{fig:setup}.
We also summarize in \cref{fig:XYdist} how the state is squeezed in the $XY$ plane.
In this setup, squeezing is performed twice by JPAs.
First, we assume that the input vacuum state $(\hat{X}_{\text{in}, m}, \hat{Y}_{\text{in}, m})$\footnote{
    The meaning of subscript $m$ is described in \cref{sec:SNR}.
} is a state of thermal photon that is sourced from the termination resistor and distributes like Gaussian in the $XY$ plane.
The first JPA called SQ in \cref{fig:setup} squeezes the vacuum state along, e.g., the $X$ direction.
When we define the squeezing parameter of the SQ JPA as $G_s$, the squeezed state $(\hat{X}_{s, m}, \hat{Y}_{s, m})$ becomes
\begin{equation}
    \hat{X}_{s, m} = \frac{1}{\sqrt{G_s}} \hat{X}_{\text{in}, m}\, , \quad \hat{Y}_{s, m} = \sqrt{G_s} \hat{Y}_{\text{in}, m}\, .
    \label{eq:squeeze_SQ}
\end{equation}
This squeezing reduces the noise $\Delta\hat{X}$.

When this squeezed state receives the signal photon from the cavity, the state is displaced in the phase of the signal photon (from the second figure to the third figure in \cref{fig:XYdist}).
Because the noise has been suppressed by a factor $1/\sqrt{G_s}$, the signal-to-noise ratio is enhanced by a factor $\sqrt{G_s}$ compared to the case without squeezing.
The second JPA called AMP squeezes the displaced state $(\hat{X}_{o, m}, \hat{Y}_{o, m})$.
This JPA amplifies the state in the $X$ direction, the opposite direction to the SQ, and we get the output state $(\hat{X}_{\text{out}, m}, \hat{Y}_{\text{out}, m})$.
Defining the squeezing parameter of the AMP JPA as $G_a$, we get
\begin{equation}
    \hat{X}_{\text{out}, m} = \sqrt{G_a} \hat{X}_{o, m}\, , \quad \hat{Y}_{\text{out}, m} = \frac{1}{\sqrt{G_a}} \hat{Y}_{o, m}\, .
    \label{eq:squeeze_AMP}
\end{equation}
Note that this second squeezing does not affect the signal-to-noise ratio because it amplifies both the signal and noise at the same time.
Instead, the AMP JPA plays a role in overwhelming the noise added by the following circuits, including the amplifier.

Technically, the direction of squeezing by JPAs is determined by the phase of the AC power input to them.
In order to give a difference to the direction of amplification by the SQ and AMP JPA, the phase shifter between the microwave generator and the SQ JPA shifts the phase of the microwaves by $\pi/2$.

\subsubsection{Homodyne measurement}

Now we need to measure the $\hat{X}$ quadrature exclusively to obtain a high signal-to-noise ratio beyond the SQL.
This is possible by using another quantum measurement technique, the homodyne measurement.
We will briefly review the theory of the homodyne measurement.
A schematic of the homodyne measurement is shown in the lower right part of \cref{fig:setup}.

First, let $\ket{\psi}$ be the signal state of our setup, i.e., the squeezed state output from the AMP JPA.
Also, in this subsection, we use the abbreviation for notation of the corresponding annihilation operator and quadratures, $\hat{a}$, $\hat{X}$, and $\hat{Y}$, representing $\hat{a}_{\mathrm{out},m}$, $\hat{X}_{\mathrm{out},m}$, and $\hat{Y}_{\mathrm{out},m}$, respectively.
The homodyne measurement requires a local oscillator that has the same mode as that of the signal photons. 
We write the annihilation operator of the local oscillator by $\hat{B}$,
 and set the initial state of the local oscillator to a coherent state
\begin{equation}
    \ket{\beta} \equiv e^{-\abs{\beta}^2/2} \sum_{n=0}^{\infty} \frac{\beta^n}{\sqrt{n!}} \ket{n}\, .
\end{equation}
Here, $\beta\equiv \abs{\beta}e^{i\theta}$ and $\ket{n}$ is the Fock state of $n$ photons.
The initial state of the total system is defined as $\ket{\Psi}\equiv\ket{\psi}\ket{\beta}$.

The signal photons and the local oscillator are split in half and mixed by a beam splitter.
As a result, we obtain two beams whose annihilation operators are
\begin{equation}
    \hat{a}' = \frac{\hat{a} - \hat{B}}{\sqrt{2}}\, , \quad \hat{B}' = \frac{\hat{a} + \hat{B}}{\sqrt{2}}\, .
\end{equation}
Next, we observe the difference $\hat{R}$ between the amplitudes of those two beams by a differential amplifier:
\begin{align}
    \hat{R} &\equiv \hat{B}^{\prime\dag} \hat{B}' - \hat{a}^{\prime\dag} \hat{a}' \notag \\
        &= \frac{\hat{a} + \hat{a}^\dag}{\sqrt{2}}\;\frac{\hat{B} + \hat{B}^\dag}{\sqrt{2}} + \frac{\hat{a} - \hat{a}^\dag}{\sqrt{2}i}\;\frac{\hat{B} - \hat{B}^\dag}{\sqrt{2}i} \, .
\end{align}
The expectation value of $\hat{R}$ is calculated as
\begin{align}
    \ev{\hat{R}}{\Psi} 
        &= \ev{\qty(\frac{\hat{a} + \hat{a}^\dag}{\sqrt{2}}\;\frac{\beta+\beta^*}{\sqrt{2}} + \frac{\hat{a} - \hat{a}^\dag}{\sqrt{2}i}\;\frac{\beta-\beta^*}{\sqrt{2}i})}{\psi} \notag \\
        &= \sqrt{2}\abs{\beta} \ev{(\hat{X}\cos\theta + \hat{Y}\sin\theta)}{\psi} \, .
\end{align}
This equation means that we can measure only one component of quadratures by observing $\hat{R}$.
For example, if $\theta=0$, we can measure only the $\hat{X}$ quadrature.
If we tune the phase $\theta$ to be the same as the phase of amplification by the AMP JPA, we can measure only the amplified quadrature.
This tuning is possible by using the same microwave generator for the AMP JPA and the local oscillator of the homodyne measurement; see \cref{fig:setup}.
Thus, the expectation value of the normalized observable $\hat{R}' \equiv \hat{R}/\sqrt{2}\abs{\beta}$ becomes
\begin{equation}
    \ev{\hat{R}'}{\Psi} = \ev{\hat{X}}{\psi}\, .
\end{equation}
Furthermore, the measurement error of the operator $\hat{R}'$ is
\begin{equation}
    \ev{(\hat{R}'-\hat{X})^2}{\Psi} = \frac{\ev{\hat{a}^\dag\hat{a}}{\psi}}{2\abs{\beta}^2}\, ,
\end{equation}
which converges to zero in the limit of $\abs{\beta}\to \infty$.
Therefore, $\hat{X}$ can be accurately measured through the homodyne measurement using the local oscillator with a large number of photons.

\section{Sensitivity}\label{sec:sensitivity}
We determine the sensitivity of our setup using a test statistic that is introduced in refs.~\cite{Foster:2017hbq,Dror:2022xpi} by developing the log-likelihood-ratio test.
Based on the discussion by refs.~\cite{Foster:2017hbq,Dror:2022xpi}, we calculate the following test statistic in order to determine the 95\% exclusion limits,
\begin{align}
    q = \frac{\tint}{\pi} \int_0^\infty \mathrm{d}\omega\, 
        \qty[\qty(1- \frac{B(\omega)}{S(\omega)+B(\omega)}) - \ln(1+\frac{S(\omega)}{B(\omega)})],
\end{align}
where $\tint$ is the experimental integration time, and $S(\omega)$ and $B(\omega)$ are the signal and noise power spectral density respectively, which are computed in \cref{sec:SNR}.
The 95\% exclusion limits are obtained by solving $q = -2.71$.

According to the calculation in \cref{sec:SNR}, we obtain
\begin{align}
    q   &\simeq -\frac{32 \geff^3  \Ncp^2 g_{ann}^4 \taum^{5/2}  \rho_a^2 \tint G_s^{1/2} Q^{3/2}}{9m_n^4 m_a^{7/2}} \notag \\
        & \simeq - 3.5 \times 10^{59} g_{ann}^4 
            \qty(\frac{\tint}{\SI{1}{min}})
            \qty(\frac{m_a}{\SI{1}{\micro eV}})^{-2}
            \qty(\frac{G_s}{10^2})^{1/2}
            \qty(\frac{M}{\SI{100}{g}})^{7/2}
            \qty(\frac{Q}{10^{6}})^{3/2}.
        \label{eq:q}
\end{align}
where $g_{ann} \equiv C_{ann} m_n / f_a$, and we have used  $\Ncp=10^{23} (M/\SI{100}{g})$, $\taum=\SI{1}{s}$, $\rho_a=\SI{0.45}{GeV/cm^{3}}$, and \cref{eq:g_eff} for $\geff$.
Note that since the cavity is placed under a low temperature $T\lesssim T_c=\order{1}\,\si{mK}$, the sensitivity does not depend on $T$ but is limited by the quantum fluctuation.
Solving $q=-2.71$, we estimate the expected exclusion limits on the axion-neutron coupling as
\begin{align}
    g_{ann} \simeq  1.7\times 10^{-15} 
            \qty(\frac{\tint}{\SI{1}{min}})^{-1/4}
            \qty(\frac{m_a}{\SI{1}{\micro eV}})^{1/2}
            \qty(\frac{G_s}{10^2})^{-1/8}
            \qty(\frac{M}{\SI{100}{g}})^{-7/8}
            \qty(\frac{Q}{10^{6}})^{-3/8} 
            .
\end{align}

In our setup, we scan the magnetic field $B_z$ and the cavity size so that the axion dark matter with mass $m_a \simeq \omega_L = \omega_{\mathrm{cavity}}$ can be searched for.
Each scan step has a sensitivity on the axion mass width $\sim 1/\tau$ around the Larmor frequency
\begin{align}
    m_a \sim \SI{0.13}{\micro eV} \left(
        \frac{B_z}{\SI{1}{T}}
    \right).
\end{align}
For simplicity, we approximate the sensitivity curve for each scan by a rectangle with width $1/\tau$ instead of using a Breit-Wigner shape.
The typical size of the cavity $L_{\mathrm{cavity}}$ is estimated by evaluating the corresponding Compton length as
\begin{align}
    L_{\mathrm{cavity}} \sim \SI{1.2}{m} \left(
        \frac{\SI{1}{\micro eV}}{m_a}
    \right).
\end{align}
The upper limit of the axion mass that can be searched by our experiment is determined by the upper limit of the magnetic field $B_z$.
We adopt $\SI{25}{T}$ as the maximum of $B_z$, which can be regarded as realistic as planned for example in CAPP25T by IBS/BNL \cite{Semertzidis:2019gkj}.\footnote{
    As a more optimistic option, $\sim \SI{45}{T}$ is also planned to be developed \cite{10.1007/978-3-030-43761-9_2}.
}

The squeezing level $G_s$ is also crucial for sensitivity estimation.
Here, we summarize the current status of the squeezing level in various experiments including the gravitational wave telescope.
The squeezing levels are usually represented in the unit of $\si{dB}$, and $x\,\si{dB}$ of squeezing corresponds to $G_s=10^{x/10}$ in our setup.
In the context of the gravitational wave detection, \SI{6}{dB} quantum noise reduction (corresponding to $G_s=10^{0.6}$) has already been reported~\cite{Lough:2020xft}, while the HAYSTAC experiment of the axion dark matter detection has achieved \SI{4}{dB} \cite{HAYSTAC:2023cam}.
Even larger values have already been achieved for the squeezed state production of light, such as \SI{8}{dB} for the microwave and the terahertz range \cite{Dassonneville:2021ntk,Kashiwazaki:2023pxy}, and \SI{15}{dB} for the megahertz range \cite{PhysRevLett.117.110801}.
It is notable, however, that a hindrance to using the squeezing state for the quantum measurement is the optical loss, which is one of the main obstacles that we have to tackle to improve the sensitivity further (see the discussion in \cref{sec:conclusion}).

\begin{figure*}[t!]
\centering
\includegraphics[width=\columnwidth]{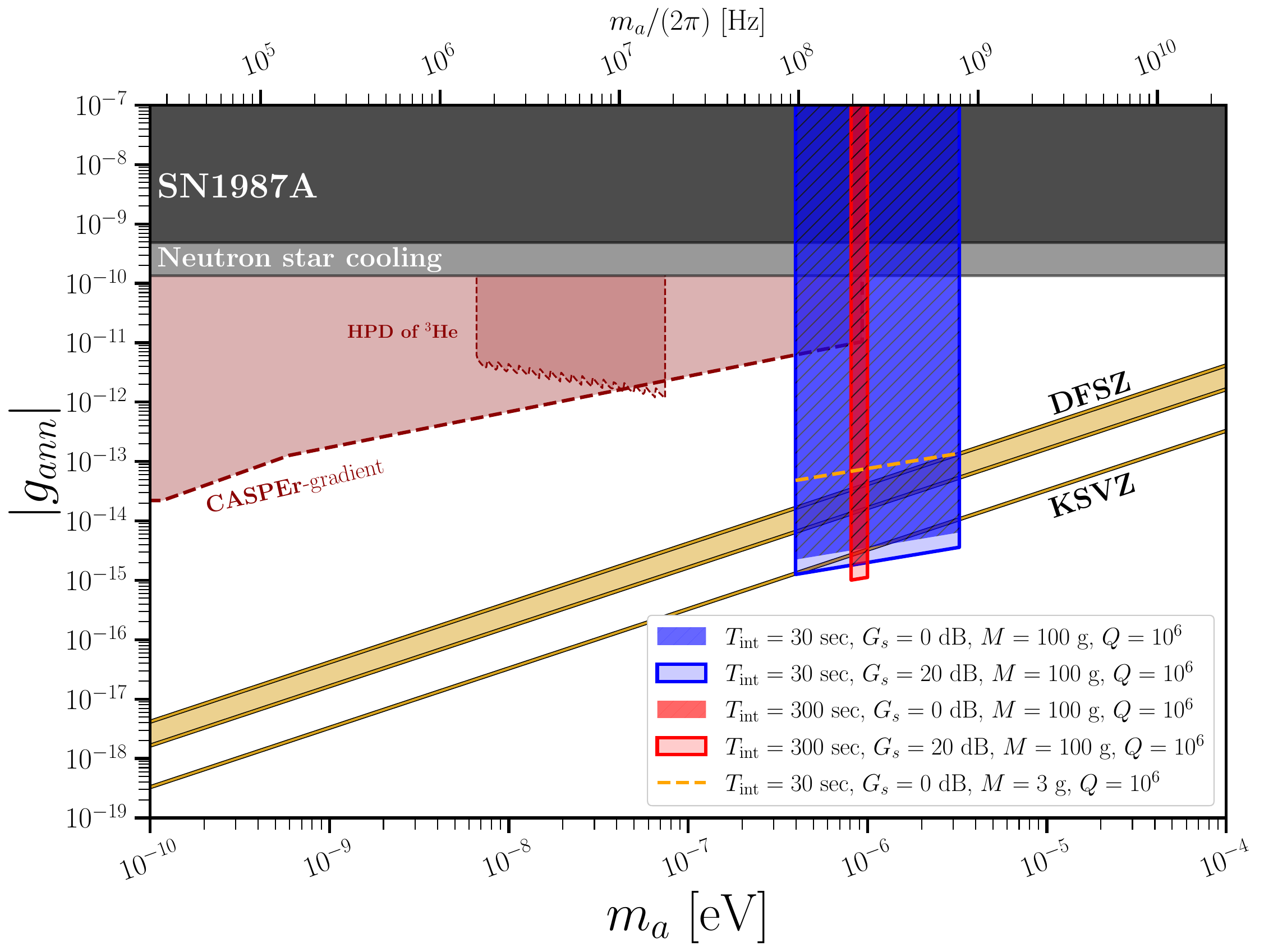}
\caption{
The 95\% exclusion limit on the axion-neutron coupling $g_{ann}$. 
We plotted the sensitivities for five setups, all of which are operated with the scan width $\tau_a^{-1}=10^{-6} m_a$ and the total integration time $T_\mathrm{tot} = \SI{2}{years}$.
The red and blue regions show the sensitivity for setups with the mass of the superfluid \he target $M=\SI{100}{g}$, the cavity's quality factor $Q=10^6$, the integration time $\tint= \SI{300}{s}$ (red) or $\SI{30}{s}$ (blue) for each scan.
The light-colored and dark-shaded regions for both colors represent the case with and without quantum measurement techniques characterized by a squeezing parameter $G_s=20 \; (0) \;\si{dB}$.
The orange dashed line shows a more realistic setup with $\tint=\SI{30}{s}$, $M=\SI{3}{g}$, $Q=10^6$, and $G_s=\SI{0}{dB}$.
The feasible upper limit of the magnetic field $B_z=\SI{25}{T}$ corresponds to the upper limits of the axion mass $m_a \simeq \SI{3.25}{\micro eV}$ for both cases.
We also plotted the region already constrained by stellar physics \cite{Carenza:2019pxu,Beznogov:2018fda},
the prospects of the CASPEr-gradient experiment \cite{JacksonKimball:2017elr} and the proposal with the homogeneous precession domain of superfluid \he \cite{Gao:2022nuq},
the prediction of the DFSZ model with $1.3\lesssim \tan\beta \lesssim 140$~\cite{Chen:2013kt,Brahmachari:1997yp}, and that of the KSVZ model.
This figure is made by using the public code~\cite{OHare:2020}. 
}
\label{fig:sensitivity}
\end{figure*}

In \cref{fig:sensitivity}, we show the 95\% exclusion limits on the axion-neutron coupling $g_{ann}$ with five benchmark setups with the total integration time fixed to $T_\mathrm{tot}=\SI{2}{years}$.
The blue regions show the sensitivities for ideal setups with $\tint=\SI{30}{s}$, $M=\SI{100}{g}$, and $Q=10^6$, with the magnetic field scanned within the range of $\SI{3.1}{T} \lesssim B_z \lesssim \SI{25}{T}$.
The red regions show other ideal setups aiming at smaller couplings of $\si{\micro eV}$ axions using longer integration time $\tint=\SI{300}{s}$, scanning the magnetic field within $\SI{6.2}{T} \lesssim B_z \lesssim \SI{7.7}{T}$.
The light-colored and dark-shaded regions for both colors represent the setups with and without quantum measurement techniques, respectively, and the former assumes the squeezing parameter of $G_s=\SI{20}{dB}$.
Here we used $Q=10^6$ as the cavity quality factor. This value is realistic compared to the state-of-the-art value of $Q=13 \times 10^7$ at a high magnetic field of $\SI{8}{T}$, which has been very recently reported by ref.~\cite{Ahn:2024pcn}.
Besides, we put a sensitivity for a more realistic setup with $\tint=\SI{30}{s}$, $M=\SI{3}{g}$, $Q=10^6$, and $G_s=\SI{0}{dB}$.
The $\SI{3}{g}$ of \he corresponds to about $\SI{1}{mol}$ of \he nuclei, a typical amount used in superfluid experiments.
All setups scan the axion mass region with a scan width corresponding to the axion width $1/\tau=1/\tau_a=10^{-6} m_a$ for each scan based on the discussion around \cref{eq:typical_time}.
We can see that our experiment can cover the predicted values of $\abs{g_{ann}}$ by the KSVZ model around $m_a\sim\si{\micro eV}$, and for the DFSZ axion, a broader range can be searched.
It can also seen that, even in a realistic setup, the parameter space can be explored beyond bounds from astronomical observations.

\section{Conclusion and discussion}
\label{sec:conclusion}

In this paper, we proposed to use the nuclear magnon modes in the ferromagnetic \Ao phase of the superfluid \he for axion DM detection.
We stressed the importance of this approach as a way to detect the axion-nucleon coupling, which is one of the most important features of the QCD axion.
As a detection method of the nuclear magnon, we proposed to use the mixing between the magnon and the cavity photon modes, which then allows us to use quantum measurement techniques such as squeezing and the homodyne measurement to enhance the DM-induced signal.
We showed the quantum mechanical description of our approach and derived the corresponding sensitivity on the axion-neutron coupling $g_{ann}$.
The result shows that our proposed approach has a sensitivity to axions with masses of about $\si{\micro eV}$, which exceeds the current best constraints by stellar physics and reaches the prediction of the KSVZ and DFSZ axion model.
Furthermore, quantum measurement techniques turned out to be useful for enhancing the sensitivity to weak signals induced by the axion DM.

In \cref{sec:sensitivity}, we have used the squeezing parameter $G_s=\SI{20}{dB}$ for benchmark setups with quantum measurement techniques.
Currently, $G_s$ has reached \SI{15}{dB}~\cite{PhysRevLett.117.110801}, and it is expected that the practical use of \SI{20}{dB} will be achieved in the near future~\cite{Kashiwazaki:2021iyd}.
However, the squeezing effect is suppressed when there is a transmission loss.
For example, when we write the transmission efficiency between the SQ JPA and the AMP JPA as $\lambda$, the effective squeezing parameter $S$ becomes
\begin{equation}
    S \simeq \qty( 1-\lambda + \frac{\lambda}{G_s})^{-1} ,
\end{equation}
as shown in ref.~\cite{Malnou:2018dxn}.
Thus, even in the limit of $G_s \to \infty$, $S$ plateaus to $(1-\lambda)^{-1}$.
Therefore, almost lossless transmission ($\lambda \simeq 1$) should be developed in order to make full use of the squeezing technology and enhance the sensitivity further. 

We may have two difficulties in our experimental setup.
One concerns the performance of the refrigerator.
We have assumed that the meter-sized cavity is cooled to about $\SI{2}{mK}$, where the \he sample becomes the \Ao phase.
Actually, there is no such refrigeration technology yet.
The CUORE experiment, which searches for neutrinoless double beta decay events, is the most successful in cooling such a large volume.
In this experiment, a cooper vessel of $\SI{1}{m^3}$ was cooled to $\SI{6}{mK}$ for 15 days using a dilution refrigerator~\cite{CUORE:2017zxr}.
In order to reach a lower temperature, we need a nuclear
demagnetization refrigerator.
This type of refrigerator is currently under rapid development, and a refrigerator that continuously reaches sub-mK has recently been developed~\cite{toda2022superconducting}.
Further development of refrigeration technology must be awaited to achieve the $\SI{2}{mK}$ cavity.

The other concerns the dynamic range of JPAs.
The tunable frequency for common JPAs is the
so-called circuit quantum electrodynamics~\cite{RevModPhys.93.025005} range of about $4\,\text{-}\,\SI{8}{GHz}$, which is larger than the Larmor frequency with the $\SI{25}{T}$ magnetic field.
We need a JPA for sub-GHz frequency to search for the axion within the mass range of our interest using the squeezing technique.
The development of sub-GHz JPAs is seen in some recent papers~\cite{vesterinen2017lumped,simbierowicz2018flux,9706259}, but the noise for them remains inferior to a SQUID amplifier.
Therefore, the implementation of JPAs in our experiment requires further refinement of the technology of sub-GHz JPAs.
Alternatively, a semi-monolithic optical parametric oscillator, which has been reported to exhibit $\SI{6.2}{dB}$ of squeezing at $\SI{2}{MHz}$~\cite{takanashi2019generation}, could be used instead of the JPA.

Finally, we comment that there are several other quantum measurement techniques that can also be applied to DM detection.
These techniques include two-mode squeezing and state-swapping interactions~\cite{Shi:2022wpf,Wurtz:2021cnm}.
For a higher frequency range, single photon counting is also a viable approach~\cite{Lamoreaux:2013koa}.  There is an attempt using superconducting qubits, which have reduced the noise to $15.7$ dB below the standard quantum limit through the repeated quantum non-demolition measurements~\cite{Dixit:2020ymh}.

\section*{Acknowledgements}

We thank Yoji Ohashi and Masahito Ueda for giving advice on the superfluid \he, and Yuta Michimura for giving information about the quantum technology of gravitational wave detectors and the likely achievements in squeezing. 
We thank Satoshi Murakawa and Tomiyoshi Haruyama for their advice on cooling techniques.
RO also thanks Satoshi Shirai and Keiichi Watanabe for pointing out a typo in the sensitivity calculation.
SC and HM are supported by the Director, Office of Science, Office of High Energy Physics of the U.S. Department of Energy under the Contract No.\ DE-AC02-05CH1123.
RO and HS are supported by Forefront Physics and Mathematics Program to Drive Transformation (FoPM), a World-leading Innovative Graduate Study (WINGS) Program, the University of Tokyo. DK and HM are supported by the Beyond AI Institute, the University of Tokyo. HM is also supported in part by the NSF grant PHY-2210390, by the JSPS Grant-in-Aid for Scientific Research JP20K03942, MEXT Grant-in-Aid for Transformative Research Areas (A) JP20H05850, JP20A203, by WPI, MEXT, Japan, and Hamamatsu Photonics, K.K. 

\appendix
\section{Statistical treatment of noise}
\label{sec:SNR}
\subsection{Formulation}

In this section, we derive the expression \cref{eq:q} of our test statistic $q$.
This quantity has been introduced as a parameter for a log-likelihood ratio test in ref.~\cite{Dror:2022xpi}.
We consider a quantum formulation of our system including the magnon and the cavity modes and apparatuses for squeezing and the homodyne measurement, and use it to evaluate the signal and the background spectral densities. 
We start with the following Hamiltonian for the cavity mode $\hat{c}$ and background modes interacting with the cavity mode:
\begin{align}
   &H_\mathrm{tot} = H_\mathrm{sys} + H_\mathrm{int} + H_\mathrm{B}\, , \\
   &H_\mathrm{sys} = \omega_L \hat{c}^\dagger \hat{c} + \omega_L \hat{d}^\dagger \hat{d} - i\frac{\Gamm}{2} \hat{d}^\dag \hat{d} + i\geff (\cavc\maga - \cava\magc), 
   \label{eq:H_sys}\\
   &H_\mathrm{int} = i \sum_{j=m,l} \sqrt{\frac{\kappa_j}{2\pi}} \int\mathrm{d}\omega\, \qty[\hat{c}^\dagger \hat{a}_j (\omega) - \hat{c}\hat{a}_j^\dagger (\omega)] + i \sqrt{\frac{\kappa_a}{2\pi}} \int\mathrm{d}\omega\, \qty[\magc \hat{a}_a (\omega) - \maga \hat{a}_a^\dagger (\omega)] , 
   \label{eq:H_int} \\
   &H_\mathrm{B} = \sum_{j=m,l,a} \int\mathrm{d}\omega\, \omega \, \hat{a}_j^\dagger (\omega) \hat{a}_j (\omega) ,
   \label{eq:H_B}
\end{align}
where $\maga$ is the annihilation operator of magnon defined by \cref{eq:magnon_spinor_relation} and $\Gamm\equiv \taum^{-1}$ is the bandwidth of magnon, and we used the rotating wave approximation.
The last term of \cref{eq:H_sys} describes the mixing of magnons and cavity modes, and the magnon field $\maga$ has been redefined in comparison to \cref{eq:polariton} for later convenience.
\Cref{eq:H_int} represents the measurement of the cavity mode, the loss of the cavity electromagnetic field, and the magnon excitation by axions as interactions with three ports: the measurement port $\hat{a}_m$, the loss port $\hat{a}_l$, and the axion port $\hat{a}_a$, respectively.
The coupling constant for the loss port $\kappa_l$ is determined with the cavity quality factor $Q$ as $\kappa_l = m_a / Q$.

In Heisenberg picture, the equations of motion for $\hat{c}(t)$, $\hat{d}(t)$, and $\hat{a}_j(\omega,t)$ are
\begin{align}
    \dv{\hat{c}(t)}{t} &= -i\omega_L \hat{c}(t) + \geff \hat{d}(t) + \sum_{j=m,l} \sqrt{\frac{\kappa_j}{2\pi}} \int \mathrm{d}\omega\, \hat{a}_j(\omega), 
    \label{eq:eom_c} \\
    \dv{\hat{d}(t)}{t} &= -i\omega_L \hat{d}(t) - \geff \hat{c}(t) - \frac{\Gamm}{2} \hat{d}(t) + \sqrt{\frac{\kappa_a}{2\pi}} \int \mathrm{d}\omega\, \hat{a}_a(\omega), 
    \label{eq:eom_d} \\
    \dv{\hat{a}_j(\omega,t)}{t} &= -i\omega\hat{a}_j(\omega,t) - \sqrt{\frac{\kappa_j}{2\pi}} 
    \left\{
    \begin{aligned}
        &\hat{c}(t) & &(j=m,l) \\
        &\hat{d}(t) & &(j=a) 
    \end{aligned}
    \right. .
    \label{eq:eom_aj}
\end{align}
The formal solution of \cref{eq:eom_aj} is written with an initial time $t_\mathrm{in}$ ($< t$) as
\begin{equation}
    \hat{a}_j(\omega,t) = e^{-i\omega(t-t_\mathrm{in})} \hat{a}_j(\omega,t_\mathrm{in})
        - \sqrt{\frac{\kappa_j}{2\pi}} \int^t_{t_\mathrm{in}} \mathrm{d}t' \; e^{-i\omega(t-t')}
        \left\{
        \begin{aligned}
            &\hat{c}(t') & &(j=m,l) \\
            &\hat{d}(t') & &(j=a) 
        \end{aligned}
        \right..
    \label{eq:sol_a_in}
\end{equation}
Substituting \cref{eq:sol_a_in} into \cref{eq:eom_c,eq:eom_d}, we get the Heisenberg--Langevin equations,
\begin{align}
    \dv{\hat{c}(t)}{t} &= -i\omega_L \hat{c}(t) - \frac{\kappa_c}{2} \hat{c}(t) + \geff \maga(t) + \sum_{j=m,l} \sqrt{\frac{\kappa_j}{2\pi}} \int \mathrm{d}\omega \; e^{-i\omega(t-t_\mathrm{in})} \hat{a}_j(\omega,t_\mathrm{in}) ,
    \label{eq:HL_c} \\
    \dv{\hat{d}(t)}{t} &= -i\omega_L \maga(t) - \frac{\kappa_d}{2} \maga(t) - \geff \hat{c}(t) + \sqrt{\frac{\kappa_a}{2\pi}} \int \mathrm{d}\omega \; e^{-i\omega(t-t_\mathrm{in})} \hat{a}_j(\omega,t_\mathrm{in}) ,
    \label{eq:HL_d} 
\end{align}
where $\kappa_c\equiv \kappa_m+\kappa_l$ and $\kappa_d\equiv \kappa_a + \Gamm$.
We define the so-called input field as
\begin{equation}
    \hat{a}_{s,j}(t) \equiv \frac{1}{\sqrt{2\pi}} \int \mathrm{d}\omega \; e^{-i\omega(t-t_\mathrm{in})} \hat{a}_j(\omega,t_\mathrm{in}) .
\end{equation}
The input field of the measurement port is shown in \cref{fig:setup}.
In terms of the input fields, \cref{eq:HL_c,eq:HL_d} are rewritten as
\begin{align}
    \dv{\hat{c}(t)}{t} &= -i\omega_L \hat{c}(t) - \frac{\kappa_c}{2} \hat{c}(t) + \geff \maga(t) + \sum_{j=m,l} \sqrt{\kappa_j} \hat{a}_{s,j} (t) ,
    \label{eq:HL_c_in} \\
    \dv{\hat{d}(t)}{t} &= -i\omega_L \maga(t) - \frac{\kappa_d}{2} \maga(t) - \geff \hat{c}(t) + \sqrt{\kappa_a} \hat{a}_{s,a} (t) .
    \label{eq:HL_d_in} 
\end{align}
We can formally solve \cref{eq:eom_aj} with another time $t_\mathrm{out}$ ($> t > t_\mathrm{in}$) as
\begin{equation}
    \hat{a}_j(\omega,t) = e^{-i\omega(t-t_\mathrm{out})} \hat{a}_j(\omega,t_\mathrm{out})
        - \sqrt{\frac{\kappa_j}{2\pi}} \int^t_{t_\mathrm{out}} \mathrm{d}t' \; e^{-i\omega(t-t')}
        \left\{
        \begin{aligned}
            &\hat{c}(t') & &(j=m,l) \\
            &\hat{d}(t') & &(j=a) 
        \end{aligned}
        \right. ,
    \label{sol_a_out}
\end{equation}
and we can also define the output field as
\begin{equation}
    \hat{a}_{o,j}(t) \equiv \frac{1}{\sqrt{2\pi}} \int \mathrm{d}\omega \; e^{-i\omega(t-t_\mathrm{out})} \hat{a}_j(\omega,t_\mathrm{out}) .
\end{equation}
Noting that the right-hand sides of \cref{eq:sol_a_in,sol_a_out} have the same form, we find the input-output relation by integrating them by $\omega$:
\begin{equation}
    \hat{a}_{o,j}(t) = \hat{a}_{s,j}(t) - \sqrt{\kappa_j} 
    \left\{
        \begin{aligned}
            &\hat{c}(t) & &(j=m,l) \\
            &\hat{d}(t) & &(j=a) 
        \end{aligned}
    \right. .
    \label{eq:input_output}
\end{equation}
The output field of the measurement port is also shown in \cref{fig:setup}.
Transforming into the rotating frame, i.e., $\hat{A}(t)\to \hat{A}(t) e^{-i\omega_L t}$ for all annihilation operators, we can eliminate the first terms in \cref{eq:HL_c_in,eq:HL_d_in}.
Thus, in the Fourier domain, we can solve \cref{eq:HL_d_in} as
\begin{equation}
    \maga(\domg) = \frac{1}{\kappa_d/2 - i\domg} \qty[-\geff \hat{c}(\domg) + \sqrt{\kappa_a} \hat{a}_{s,a}(\domg)] ,
\end{equation}
where $\domg\equiv \omega-\omega_L$.
Substituting this into \cref{eq:HL_c_in} and using the input-output relation for the measurement port, we get
\begin{equation}
    \hat{a}_{o,m}(\domg) = \sum_{j=m,l,a} \chi_{j}(\domg) \hat{a}_{s,j}(\domg) 
    \label{eq:chi},
\end{equation}
where
\begin{align}
    \chi_{j}(\domg) &= \delta_{mj} - \qty(\frac{\kappa_c}{2} + \frac{\geff^2}{\kappa_d/2 - i\domg} - i\domg)^{-1} \notag \\
        &\qquad \times \sqrt{\kappa_m\kappa_j}
        \left\{
        \begin{aligned}
            &1 && (j=m,l) \\
            &\frac{\geff}{\kappa_d/2-i\domg} && (j=a)
        \end{aligned}
        \right. .
\end{align}
Note that the susceptibility $\chi_j(\domg)$ satisfies $\chi_j^*(-\Delta\omega)={\chi_j}(\Delta\omega)$.

We move to the quadrature basis by
\begin{align}
    \begin{pmatrix}
        \hat{X}(\Delta\omega)\\
        \hat{Y}(\Delta\omega)
    \end{pmatrix}
    =\frac{1}{\sqrt{2}}
    \begin{pmatrix}
        1 && 1\\
        -i && i
    \end{pmatrix}
    \begin{pmatrix}
        \hat{a}(\Delta\omega)\\
        \hat{a}^\dag(-\Delta\omega)
    \end{pmatrix}
    \equiv {\bm P}
    \begin{pmatrix}
        \hat{a}(\Delta\omega)\\
        \hat{a}^\dag(-\Delta\omega)
    \end{pmatrix} .
\end{align}
We would like to relate the output operator $\hat{X}_{\mathrm{out},m}$, which goes out from the AMP JPA, with the input operator $\hat{X}_{\mathrm{in},m}$, which comes into the SQ JPA, by an SSR/cavity susceptibility $\Xi_j$ as $\hat{X}_{\mathrm{out},m} =  \sum_j \Xi_j \hat{X}_{\mathrm{in},j}$.
First, the SQ JPA squeezes $\hat{X}_{\mathrm{in},m}$:
\begin{align}
    \vec{\hat{X}}_{s,m}(\domg) = \frac{1}{\sqrt{G_s}} \vec{\hat{X}}_{\text{in},m} (\domg),
\end{align}
while leaving the operators at other ports unaffected.
The SQ JPA amplifies the other quadrature at the measurement port $\hat{Y}_{\text{in},m}$ by $\sqrt{G_s}$, but we do not track the $\hat{Y}$ quadratures because we will only measure the $\hat{X}$ quadrature.
Using \cref{eq:chi}, we find that the susceptibility in the quadrature basis is the same as that in the original basis $\chi_j(\domg)$:
\begin{align}
    \begin{pmatrix}
        \hat{X}_{o,m} (\domg)\\
        \hat{Y}_{o,m} (\domg)
    \end{pmatrix}
    & = {\bm P}
    \begin{pmatrix}
        \sum_j \chi_j(\Delta\omega) && 0\\
        0 && \sum_j \chi_j^*(-\Delta\omega)
    \end{pmatrix}
    {\bm P^{-1}}
    \begin{pmatrix}
        \hat{X}_{s,j} (\domg)\\
        \hat{Y}_{s,j} (\domg)
    \end{pmatrix}
    \notag \\
    &= \sum_j 
    \begin{pmatrix}
        \chi_j(\Delta\omega) && 0\\
        0 && \chi_j(\Delta\omega)
    \end{pmatrix}
    \begin{pmatrix}
        \hat{X}_{s,j} (\domg)\\
        \hat{Y}_{s,j} (\domg)
    \end{pmatrix}
    .
\end{align}
Hence, $\hat{X}_{o,m} (\domg)= \sum_j \chi_j(\Delta\omega) \hat{X}_{s,j} (\domg)$.
Finally, the AMP JPA performs amplification with a squeezing parameter $G_a$ as
\begin{align}
    \hat{X}_{\mathrm{out},m} (\domg)
    = \sqrt{G_a} \hat{X}_{o,m} (\domg).
\end{align}
As a result, we get the SSR/cavity susceptibility,
\begin{align}
    \Xi_j (\domg)
    = \left\{ 
    \begin{aligned}
        &\sqrt{\frac{G_a}{G_s}} \chi_j(\domg) && (j=m) \\
        &\sqrt{G_a} \chi_j(\domg) && (j=l,a)
    \end{aligned}
    \right. ,
\end{align}
and accordingly,
\begin{align}
    \hat{X}_{\mathrm{out},m}(\domg) 
    &= \sum_{j=m,l,a} \Xi_j (\domg) \hat{X}_{\mathrm{in},j}(\domg) \notag \\
    &= \sqrt{G_a} \qty[\frac{\chi_{m}(\domg)}{\sqrt{G_s}} \hat{X}_{\mathrm{in},m}(\domg) + \sum_{j=l,a} \chi_{j}(\domg) \hat{X}_{\mathrm{in},j}(\domg)].
    \label{eq:Xout}
\end{align}

Next, we calculate the output power spectral density (PSD), $P(\domg)$.
\Cref{eq:Xout} leads to
\begin{align}
    P(\domg) &\equiv \frac{1}{\tint} \expval{\hat{X}^\dag_{\mathrm{out},m}(\domg) \hat{X}_{\mathrm{out},m}(\domg)} \notag \\
    &= \frac{G_a}{\tint} \Biggl[\frac{\abs{\chi_{mm}(\domg)}^2}{G_s} \expval{\hat{X}_{\mathrm{in},m}^\dag(\domg)\hat{X}_{\mathrm{in},m}(\domg)} \notag \\
    &\qquad + \sum_{j=l,a} \abs{\chi_{mj}(\domg)}^2 \expval{\hat{X}_{\mathrm{in},j}^\dag(\domg)\hat{X}_{\mathrm{in},j}(\domg)} \Biggr],
\end{align}
where $\tint$ is the integration time for each scan.
The input spectral densities are
\begin{align}
    &\frac{1}{\tint}\expval{\hat{X}_{\mathrm{in},m}^\dag(\domg)\hat{X}_{\mathrm{in},m}(\domg)} = \frac{1}{\tint}\expval{\hat{X}_{\mathrm{in},l}^\dag(\domg)\hat{X}_{\mathrm{in},l}(\domg)} = n_T + \frac{1}{2}, \\
    &\frac{1}{\tint}\expval{\hat{X}_{\mathrm{in},a}^\dag(\domg)\hat{X}_{\mathrm{in},a}(\domg)} = n_a + \frac{1}{2},
\end{align}
where $n_T$ and $n_a$ are the numbers of the input thermal photon and the axion per unit time per unit bandwidth, respectively. 
We assumed that the thermal noise dominates the input noise for the measurement port and the loss port.
Note that $n_T$ and $1/2$s in the spectral density matrix correspond to the thermal and the quantum noises, respectively.
We also assume $n_T \ll 1/2$ since our experiment is operated under a low temperature $T< T_c =\order{1}\;\si{mK}$, which is lower than the SQL temperature $T_\mathrm{SQL} = m_a/k_B = \order{10}\;\si{mK}$.
Decomposing the PSD into the signal and the noise part, we get the signal and the noise spectral densities $S(\domg)$ and $B(\domg)$,
\begin{align}
    S(\domg) &= \frac{G_a}{\tint} \abs{\frac{\kappa_c}{2} + \frac{\geff^2}{\kappa_d/2 - i\domg} - i\domg}^{-2} \frac{\geff^2 \kappa_m}{(\kappa_d/2)^2+(\domg)^2}  \kappa_a n_a, \\
    B(\domg) &= \frac{G_a}{2\tint} \abs{\frac{\kappa_c}{2} + \frac{\geff^2}{\kappa_d/2 - i\domg} - i\domg}^{-2} \notag \\
        &\quad \times \Biggl[
            \frac{1}{G_s} \qty{\qty(\frac{-\kappa_m + \kappa_l}{2}+\frac{\geff^2}{(\kappa_d/2)^2+(\domg)^2} \frac{\kappa_d}{2})^2 + \qty(\frac{\geff^2}{(\kappa_d/2)^2+(\domg)^2}-1)^2 (\domg)^2} \notag \\
        &\qquad \qquad + \kappa_m\kappa_l + \frac{\geff^2 \kappa_m\kappa_a}{(\kappa_d/2)^2+(\domg)^2}
        \Biggr].
\end{align}

\subsection{Creation rate of magnons}
Let us compute the creation rate of magnons in order to estimate $\kappa_a n_a$.
We start with the equation of motion for the magnon operator,
\begin{equation}
    \dv{\maga(t)}{t} = -i\omega_L\maga(t) - \frac{\Gamm}{2} \maga(t) - i \frac{C_{ann}}{f_a} \sqrt{\rho_a \Ncp} v_a^+(t) \sin[\omega_L t + \varphi(t)],
\end{equation}
where the last term comes from the axion-magnon interaction derived in \cref{eq:axion-magnon-int}.
Under the assumption $\maga(0) = 0$, the formal solution is
\begin{equation}
    \maga (t) = -i \frac{C_{ann}}{f_a} \sqrt{\rho_a\Ncp} \int_0^t \mathrm{d}t'\; e^{(-i\omega_L-\Gamm/2) (t-t')} v_a^+(t') \sin[\omega_L t' + \varphi(t')].
\end{equation}
It is convenient to introduce the autocorrelation function $C(t,t') \equiv \expval{\magc(t)\maga(t')}$, where the expectation value is taken for the stochastic values: the axion velocity $v_a(t)$ and the phase $\varphi(t)$.
For $t,t' \gg \tau_a$, where $\tau_a$ is the axion coherence time $\tau_a\simeq (m_av_a^2)^{-1}$, we can compute $C(t,t')$ as
\begin{align}
    C(t,t')
    &= \qty(\frac{C_{ann}}{f_a})^2 \rho_a\Ncp 
        \int_0^t \mathrm{d}\bar{t} \int_0^{t'} \mathrm{d}\bar{t}' \; 
        e^{(+i\omega_L-\Gamm/2) (t-\bar{t})} e^{(-i\omega_L-\Gamm/2) (t'-\bar{t}')} \notag \\
    &\qquad \times \expval{v_a^-(\bar{t}) v_a^+(\bar{t}') 
        \sin[\omega_L\bar{t} + \varphi(\bar{t})] \sin[\omega_L\bar{t}' + \varphi(\bar{t}')]} \notag \\
    &\simeq \qty(\frac{C_{ann}}{f_a})^2 \rho_a\Ncp 
        \int_0^t \mathrm{d}\bar{t} \int_0^{t'} \mathrm{d}\bar{t}' \;
        e^{(+i\omega_L-\Gamm/2) (t-\bar{t})} e^{(-i\omega_L-\Gamm/2) (t'-\bar{t}')} \notag \\
    &\qquad \times \frac{1}{3} v_a^2 \cos[\omega_L(\bar{t} - \bar{t}')] \Theta(\tau_a - \abs{\bar{t}-\bar{t}'}) \notag \\
    &\simeq \frac{2}{3} \qty(\frac{C_{ann}}{f_a})^2 \rho_a\Ncp v_a^2
        \tau_a e^{i\omega_L (t-t')} e^{-(\Gamm/2) (t+t')} 
        \Gamm^{-1} \qty[e^{\Gamm \min[t,t']} - 1].
\end{align}
In order to get the second line, we used an assumption that stochastic quantities do not correlate unless $\abs{\bar{t}-\bar{t}'} < \tau_a$.
The spectral density is obtained by Fourier-transforming $C(t,t')$,
\begin{align}
    \frac{1}{\tint} \expval{\magc(\omega)\maga(\omega)} 
        &= \frac{1}{\tint} \int_0^{\tint} \mathrm{d}t \int_0^{\tint} \mathrm{d}t' \;
            e^{-i\omega(t-t')} C(t,t') \notag \\
        &\simeq \frac{8}{3} \qty(\frac{C_{ann}}{f_a})^2 \rho_a\Ncp v_a^2
        \tau_a \frac{1}{\Gamm^2 + 4\domg^2}.
        \label{eq:corr_Fourier}
\end{align}
Here, we used $\tint \gg \taum = \Gamm^{-1}$.

Next, we will estimate $\kappa_a n_a$.
The solution of \cref{eq:HL_d_in} in the Fourier domain with $\hat{c}=0$ leads to
\begin{equation}
    \hat{a}_{s,a}(\domg) = \frac{\kappa_d/2 - i\domg}{\sqrt{\kappa_a}} \maga(\domg) 
        \simeq \frac{\Gamm/2 - i\domg}{\sqrt{\kappa_a}} \maga(\domg),
\end{equation}
where we assumed $\kappa_a \ll \Gamm$.
Thus, $n_a$ is estimated as 
\begin{align}
    n_a &= \frac{1}{\tint} \expval{\hat{a}^\dag_{\mathrm{in},a} (\domg) \hat{a}_{\mathrm{in},a}(\domg)} \notag \\
        &\simeq \frac{(\Gamm/2)^2 + \domg^2}{\kappa_a} \frac{1}{\tint} \expval{\magc(\domg)\maga(\domg)} \notag \\
        &= \frac{2}{3\kappa_a} \qty(\frac{C_{ann}}{f_a})^2 \rho_a\Ncp v_a^2
        \tau_a ,
\end{align}
and hence,
\begin{equation}
    \kappa_a n_a \simeq \frac{2}{3} \qty(\frac{C_{ann}}{f_a})^2 \rho_a\Ncp v_a^2
        \tau_a .
    \label{eq:kana}
\end{equation}
Considering that $n_a$ has a bandwidth $\Delta_a \simeq m_av_a^2$, which reflects the axion coherence, we should modify \cref{eq:kana} as
\begin{equation}
    \kappa_a n_a \simeq \frac{2}{3} \qty(\frac{C_{ann}}{f_a})^2 \rho_a\Ncp v_a^2
        \tau_a \Theta(\Delta_a/2 - \abs{\domg}).
\end{equation}

\subsection{Test statistic}

In order to determine the 95\% exclusion limit, we introduce a log-likelihood ratio test statistic $q$ \cite{Foster:2017hbq,Dror:2022xpi}. 
It is computed in the limits of $\tint \gg \{\taum, \tau_a\}$ and $S(\domg)\ll B(\domg)$ as 
\begin{align}
    q &\simeq \frac{\tint}{\pi} \int_0^\infty \mathrm{d}\omega\, 
        \qty[\qty(1- \frac{B(\omega)}{S(\omega)+B(\omega)}) - \ln(1+\frac{S(\omega)}{B(\omega)})]  \notag \\
    &\simeq -\frac{\tint}{2\pi} \int_0^\infty \mathrm{d}(\domg)\; \qty(\frac{S(\domg)}{B(\domg)})^2 .
\end{align}
When we assume $\kappa_a \ll \{\geff, \Gamm, \kappa_m, \kappa_l\}$, we can approximate $q$ as
\begin{align}
    q 
    &\simeq -\frac{8 \geff^4 \kappa_m^2 \Ncp^2 g_{ann}^4 \rho_a^2 v_a^4 \tau_a^2 \tint}{9\pi m_n^4} G_s^2 \notag \\
    &\qquad \times \int_{-\Delta_a/2}^{\Delta_a/2} \mathrm{d}(\domg)\; \Biggl[
           (\domg)^4 + \qty{\qty(\frac{-\kappa_m + \kappa_l}{2})^2 + \qty(\frac{\Gamm}{2})^2 - 2\geff^2 + G_s \kappa_m \kappa_l} {(\domg)}^2 \notag \\
    &\qquad \qquad + \qty(\frac{-\kappa_m + \kappa_l}{2}\frac{\Gamm}{2} + \geff^2)^2
            + G_s \kappa_m\kappa_l \qty(\frac{\Gamm}{2})^2
        \Biggr]^{-2} .
    \label{eq:q_original}
\end{align}
The 95\% exclusion limit corresponds to the point $q\simeq -2.71$.

The parameter $\kappa_m$ determines the speed of the signal readout, and we can choose the optimal coupling $\kappa_m$ so that the size of the test statistic $|q|$ is maximized.
For this purpose, we discuss the maximization of the following integral,
\begin{align}
    I(\kappa_m) &= \kappa_m^2 \int_{-\Delta_a/2}^{\Delta_a/2} \mathrm{d}(\domg)\; \Biggl[
           (\domg)^4 \notag \\
        &\qquad + \qty{\qty(\frac{-\kappa_m + \kappa_l}{2})^2 + \qty(\frac{\Gamm}{2})^2 - 2\geff^2 + G_s \kappa_m \kappa_l} {(\domg)}^2 \notag \\
        &\qquad + \qty(\frac{-\kappa_m + \kappa_l}{2}\frac{\Gamm}{2} + \geff^2)^2
        + G_s \kappa_m\kappa_l \qty(\frac{\Gamm}{2})^2
    \Biggr]^{-2}  \notag \\
    &\equiv \kappa_m^2 \int_{-\Delta_a/2}^{\Delta_a/2} \mathrm{d}(\domg)\; \qty[(\domg)^4 + \xi(\kappa_m) (\domg)^2 + \zeta(\kappa_m)]^{-2} .
    \label{eq:Integral}
\end{align}
The integral \eqref{eq:Integral} is enhanced when $\kappa_m$ is fine-tuned to $\kappa_m \simeq \kappa_l + 4\geff^2/\Gamm \equiv \kappa_m^*$.
With such fine-tuning, the width of the integrand of \cref{eq:Integral} becomes sufficiently smaller than $\Delta_a$, and hence the integration range is approximated as $-\infty$ to $+\infty$:
\begin{align}
    I(\kappa_m^*) 
        &\simeq (\kappa_m^*)^2 \int_{-\infty}^{\infty} \mathrm{d}(\domg)\; \frac{1}{[\domg^4 + \xi(\kappa_m^*) \domg^2 + \zeta(\kappa_m^*)]^2} \notag \\
        &= (\kappa_m^*)^2 \frac{\pi}{2\sqrt{\xi(\kappa_m^*) \zeta(\kappa_m^*)^3}} .
\end{align}
Therefore, substituting $\kappa_m = \kappa_m^*$ into \cref{eq:q_original}, we obtain
\begin{align}
    q   &\simeq -\frac{64 \geff^4  \Ncp^2 g_{ann}^4 \rho_a^2 v_a^4 \tau_a^2 \tint G_s^{1/2}}{9m_n^4 \Gamm^5} \notag \\
        & \qquad \times \qty(\gamma_\mathrm{eff}^2 + \gamma_l)^2 
        \qty[\qty(\gamma_\mathrm{eff}^2 - 1)^2 + 4 G_s \gamma_l \qty(\gamma_\mathrm{eff}^2 + \gamma_l)]^{-1/2} 
        \qty[\gamma_l \qty(\gamma_\mathrm{eff}^2 + \gamma_l)]^{-3/2},
\end{align}
where we have defined the following dimensionless variables,
\begin{align}
    \gamma_\mathrm{eff} &\equiv \frac{2\geff}{\Gamm} = \order{10^6} , \\
    \gamma_l &\equiv \frac{\kappa_l}{\Gamm} = \order{10^3} \qty(\frac{m_a}{\si{\micro eV}}) \qty(\frac{Q}{10^6})^{-1} .
\end{align}
Since $\gamma_\mathrm{eff}^2$ is sufficiently larger than $\gamma_l$ for $m_a \sim \si{\micro eV}$, we can approximate $q$ as
\begin{align}
    q   &\simeq -\frac{64 \geff^4  \Ncp^2 g_{ann}^4 \rho_a^2 v_a^4 \tau_a^2 \tint G_s^2}{9m_n^4 \Gamm^{1/2}} \gamma_\mathrm{eff}^{-1} \gamma_l^{-3/2} \notag \\
        &= -\frac{32 \geff^3  \Ncp^2 g_{ann}^4 \rho_a^2 \tint G_s^{1/2} Q^{3/2}}{9m_n^4 \Gamm^{5/2} m_a^{7/2}} \notag \\
        &\simeq - 3.5 \times 10^{59} g_{ann}^4 
            \qty(\frac{\tint}{\SI{1}{min}})
            \qty(\frac{m_a}{\SI{1}{\micro eV}})^{-2}
            \qty(\frac{G_s}{10^2})^{1/2}
            \qty(\frac{M}{\SI{100}{g}})^{7/2}
            \qty(\frac{Q}{10^{6}})^{3/2}.
        \label{eq:q_app}
\end{align}
Here we have used $\geff/2\pi=\SI{0.3}{MHz}$, $\Ncp=10^{23} (M/\SI{100}{g})$, $\Gamm=\SI{1}{Hz}$, and $\rho_a=\SI{0.45}{GeV/cm^{3}}$ to get the last line.

\section{Josephson parametric amplifier (JPA)}
\label{sec:JPA}

Here, we will review Josephson parametric amplifier (JPA). 

\subsection{Effective description}
Let us imagine that the circuit model has two junctions at $x_1$ and $x_2$.
The effective Hamiltonian is given by 
\begin{align}
    H &=\int \mathrm{d}^3x \left[\frac{1}{2m}|\vec{D}\Psi_1|^2+V(\Psi_1)
    +\frac{1}{2m}|\vec{D}\Psi_2|^2+V(\Psi_2)\right] \\
    &\qquad +\alpha\Psi_2^*\Psi_1 \left[\delta(x-x_1)+\delta(x-x_2)\right]+\text{h.c.}, 
\end{align}
where $x_1,x_2$ represent the places of the junctions, and
$\Psi_1, \Psi_2$ are the wave functions of Cooper pairs, where $\Psi_1(x)$ lives within the interval $x_1 < x < x_2$, while $\Psi_2(x)$ within $x_2 < x < x_1$ (note that this is a loop).
In the case of a circuit without any junctions, energy minimization requires 
\begin{align}
    \vec{0} = \vec{D}\Psi = \vec{\nabla} \Psi -i2e\vec{A}\Psi
    =i|\Psi|(\vec{\nabla}\theta-2e\vec{A}),
\end{align}
where $\theta$ is the phase of $\Psi$ and $\vec{A}$ is the photon field.
Integrating along the circuit, we get
\begin{align}
    2\pi n=\oint \vec{\nabla}\theta\cdot \mathrm{d}\vec{r}=\oint 2e\vec{A}\cdot \mathrm{d}\vec{r} =2e\Phi,
\end{align}
with an integer $n$.
We used the single-valuedness of the wave function for the left equation, and $\Phi$ is the magnetic flux. 
For the Josephson junction, however, $\vec{\nabla} \theta_{1,2}$ need to be treated independently and hence the magnetic flux does not need to be quantized,
\begin{align}
    2e\Phi = \oint \vec{\nabla}\theta=\int_{x_1}^{x_2} \vec{\nabla} \theta_1
    + \int_{x_2}^{x_1} \vec{\nabla} \theta_2 
    = \theta_1(x_2)-\theta_1(x_1) + \theta_2(x_1) - \theta_{2}(x_2).
\end{align}

For our purposes, we are not interested in the dynamics in the bulk of the superconductor where all excitations are gapped but rather only in the junctions where the phase degrees of freedom can have much smaller excitation energies. Noting the canonical commutation relation
\begin{align}
    [ \Psi (x), \Psi^\dagger(y)] = \delta(x-y),
\end{align}
and rewriting it as $\Psi(x) = \sqrt{N(x)} e^{i\theta}(x)$, we can derive\footnote{
Considering the periodicity of $\theta$, the right commutation relation of these variables is
\begin{align*}
    [e^{i\theta(x)},\,N(y)]=e^{i\theta(x)}
    \delta(x-y).
\end{align*}
}
\begin{align}
    [\theta(x), N(y) ] = i \delta(x-y).
\end{align}
In addition, we are only interested in the phase differences across the junction. Therefore we reduce the degrees of freedom down to $\vartheta_1 = \theta_2(x_1) - \theta_1(x_1)$ and $\vartheta_2 = \theta_1(x_2) - \theta_2(x_2)$ subject to the constraint $\vartheta_1 + \vartheta_2 = 2e\Phi$. 

On the other hand, across the junctions we expect a capacitance $C$ so that the Hamiltonian contains
\begin{align}
    \frac{Q(x_{1,2})^2}{2C} = \frac{(2e)^2}{2C} n_{1,2}^2 ,
\end{align}
where we defined $n_{1}=\frac{1}{2}N_{2}(x_{1})-\frac{1}{2}N_{1}(x_{1})$ and $n_{2}=\frac{1}{2}N_{1}(x_{2}) -\frac{1}{2}N_{2}(x_{2})$.
Here we made a simplification that the capacitance is the same for both junctions.
Combining them together, we find the simplified Hamiltonian
\begin{align}
    H= \frac{2e^2}{C} (n_{1}^2 + n_{2}^2) 
    + 2\alpha (\sqrt{N_{1}(x_{1})N_{2}(x_{1})} \cos\vartheta_1 + \sqrt{N_{1}(x_{2})N_{2}(x_{2})} \cos\vartheta_2).
\end{align}
We define $\vartheta \equiv (\vartheta_1 - \vartheta_2)/2$ and its canonical conjugate $n=n_1-n_2$. 
Assuming that two Josephson energies are same, $2\alpha\sqrt{N_{1}(x_{1})N_{2}(x_{1})}= 2\alpha\sqrt{N_{1}(x_{2})N_{2}(x_{2})}= -E_{J}$, we get
\begin{align}
    H = \frac{e^2}{C} n^2 
    -2E_{J}\cos(e\Phi)\cos(\vartheta).\label{eq:C.10}
\end{align}
Here we neglected the term proportional to $(n_{1}+n_{2})^{2}$ since the value of $n_{1}+n_{2}$ is conserved.
\subsection{Flux-driven  Josephson parametric amplifier}
\begin{figure}
    \centering
    \includegraphics[width=10cm]{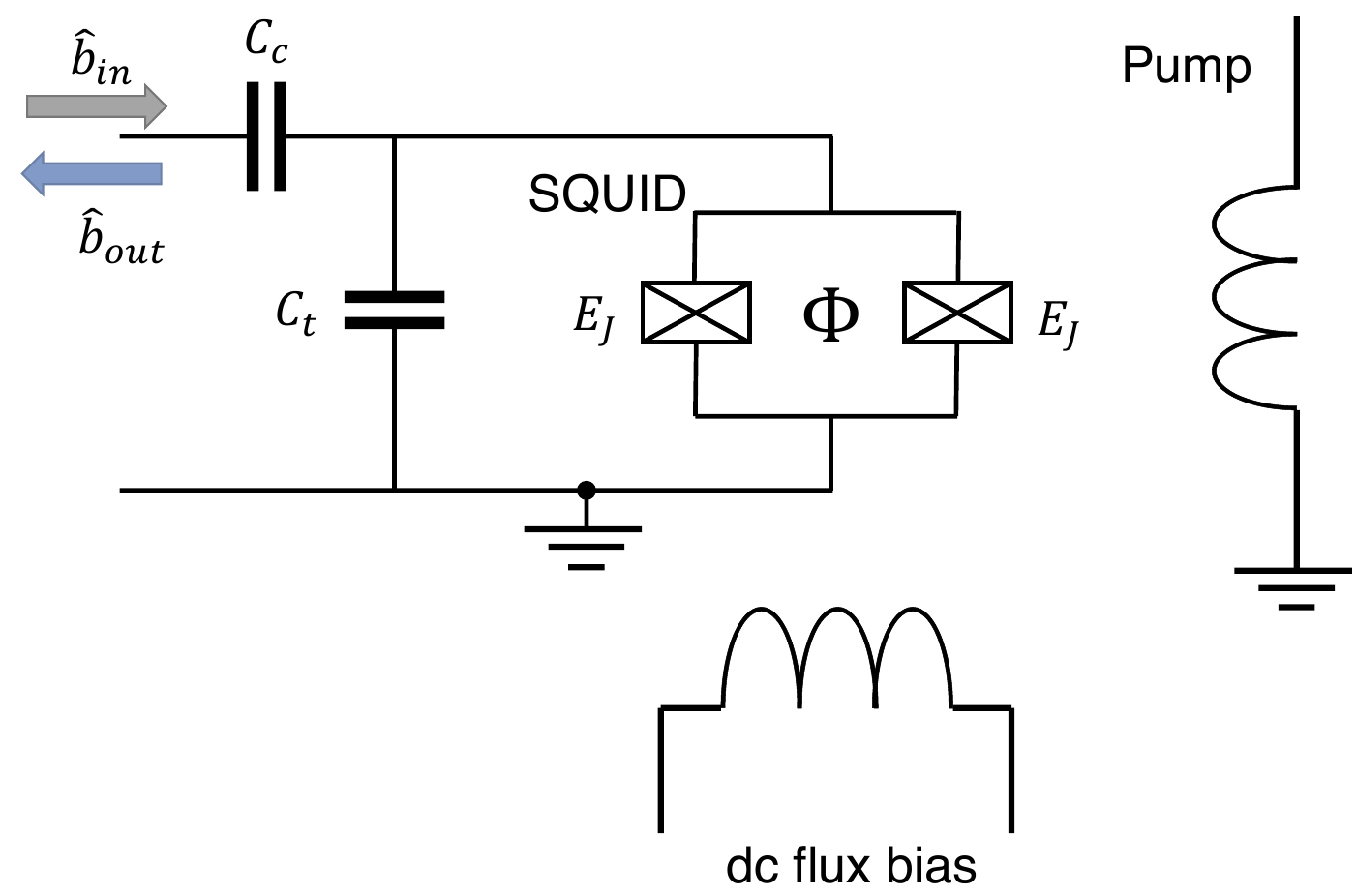}
    \caption{Schematic of a flux-driven Josephson parametric amplifier (FJPA). It consists of a SQUID biased by the external ﬂux $\Phi_{\text{ext}}=\Phi_{\text{DC}}+\Phi_{\text{AC}}\cos(\alpha\omega_{c}  t)$ and a shunting capacitance $C_{t}$, and is connected to the input/output port. We assume two junctions in the SQUID have the same Josephson energies $E_{J}$.}
    \label{fig:JPA_circuit}
\end{figure}
In the following, we explain how the flux-driven Josephson parametric amplifiers (FJPA) work in squeezing and amplifying the signal. We consider the simplified theoretical model of the FJPA (\cref{fig:JPA_circuit}). The FJPA consists of a SQUID biased by the external ﬂux $\Phi_{\text{ext}}$ and a shunting capacitance $C_{t}$ and is connected to the input/output port.\\
 Similarly to \cref{eq:C.10}, The Hamiltonian describing the resonator part of FJPA is 
\begin{align}
    H_{\text{sys}}&=\frac{(2en)^{2}}{2C_{t}}-E_{J}\cos\vartheta_{1}-E_{J}\cos\vartheta_{2} \notag \\
        &=4E_{C}n^{2}-E_{J}^{\text{eff}}(\Phi_{\text{ext}})\cos\vartheta,
\end{align}
where $E_{C}=e^{2}/(2C_{t})$, $E_{J}^{\text{eff}}(\Phi_{\text{ext}})=2E_{J}\cos(e\Phi_{\text{ext}})
=2E_{J}\cos(\pi\Phi_{\text{ext}}/\Phi_{0})$. We set the DC part of $\Phi_{\text{ext}}$ to the quarter of 
magnetic flux quantum, i.e., bias the ampliﬁer at $\Phi_{DC}=\Phi_{0}/4$.
In the absence of a pump, 
\begin{align}
    H_{\text{sys}}&=4E_{C}n^{2}-\sqrt{2}E_{J}\cos\vartheta.
\end{align}
Expanding $\cos\vartheta$ to order $\vartheta^{2}$, we can write the Hamiltonian by the ladder operator.
\begin{align}
    H_{\text{sys}}&=\omega_{c}\hat{a}^{\dagger}\hat{a},
\end{align}
where
\begin{align}
    \vartheta&=\left(
    \frac{\sqrt{2}E_{C}}{E_{J}}
    \right)^{1/4}(\hat{a}^{\dagger}+\hat{a}), \\
    n &=\frac{i}{2}\left(
    \frac{E_{J}}{\sqrt{2}E_{C}}
    \right)^{1/4}(\hat{a}^{\dagger}-\hat{a}), \\ 
    \omega_{c} &= 2\sqrt{2\sqrt{2}E_{C}E_{J}}.
\end{align}
Next, we consider including the AC part of the external field $\Phi_{\text{ext}}$ due to the pumping
\begin{align}
    \Phi_{\text{ext}}&=\Phi_{\text{DC}}+\Phi_{\text{AC}}\cos(\alpha\omega_{c}t) .
\end{align}
We set the AC part of $\Phi_{\text{ext}}$ smaller than the DC part $\Phi_{\text{AC}}\ll\Phi_{\text{DC}}$, and evaluate $E_{J}^{\text{eff}}(\Phi_{\text{ext}})$ as 
\begin{align}
    E_{J}^{\text{eff}}(\Phi_{\text{ext}})
    &\simeq E_{J}^{\text{eff}}(\Phi_{\text{DC}})
    +\left.\frac{\partial E_{J}^{\text{eff}}
    (\Phi)
    }{\partial \Phi}\right|_{\Phi=\Phi_{\text{DC}}}
    \Phi_{\text{AC}}\cos(\alpha\omega_{c}t) \notag \\
    &=\sqrt{2}E_{J}-\left(\frac{\pi \Phi_{\text{AC}}}{\Phi_{0}}\right)\sqrt{2}E_{J}\cos(\alpha \omega_{c}t) .
\end{align}
Thus, Hamiltonian $H_{\text{sys}}$ becomes
\begin{align}
    H_{\text{sys}}&\simeq  \omega_{c}\hat{a}^{\dagger}\hat{a}+\mu_{r}
    \cos(\alpha\omega_{c}t)(\hat{a}^{\dagger}+\hat{a})^{2},
\end{align}
where $\mu_{r}=\frac{\pi \Phi_{\text{AC}}}{\Phi_{0}}\left(\frac{E_{C}E_{J}}{\sqrt{2}}\right)^{\frac{1}{2}}$. We focus on the parametric amplifier region ($\alpha\simeq2$). Applying rotating wave approximation, we can estimate $H_{\text{sys}}$ as
\begin{align}
    H_{\text{sys}} &\simeq  \omega_{c}\hat{a}^{\dagger}\hat{a}+
    \frac{\mu_{r}}{2}e^{i\alpha\omega_{c}t}\hat{a}^{2}+
    \frac{\mu_{r}}{2}e^{-i\alpha\omega_{c}t}\hat{a}^{\dagger 2}.
\end{align}

\begin{figure}
    \centering
    \includegraphics[width=10.0cm]{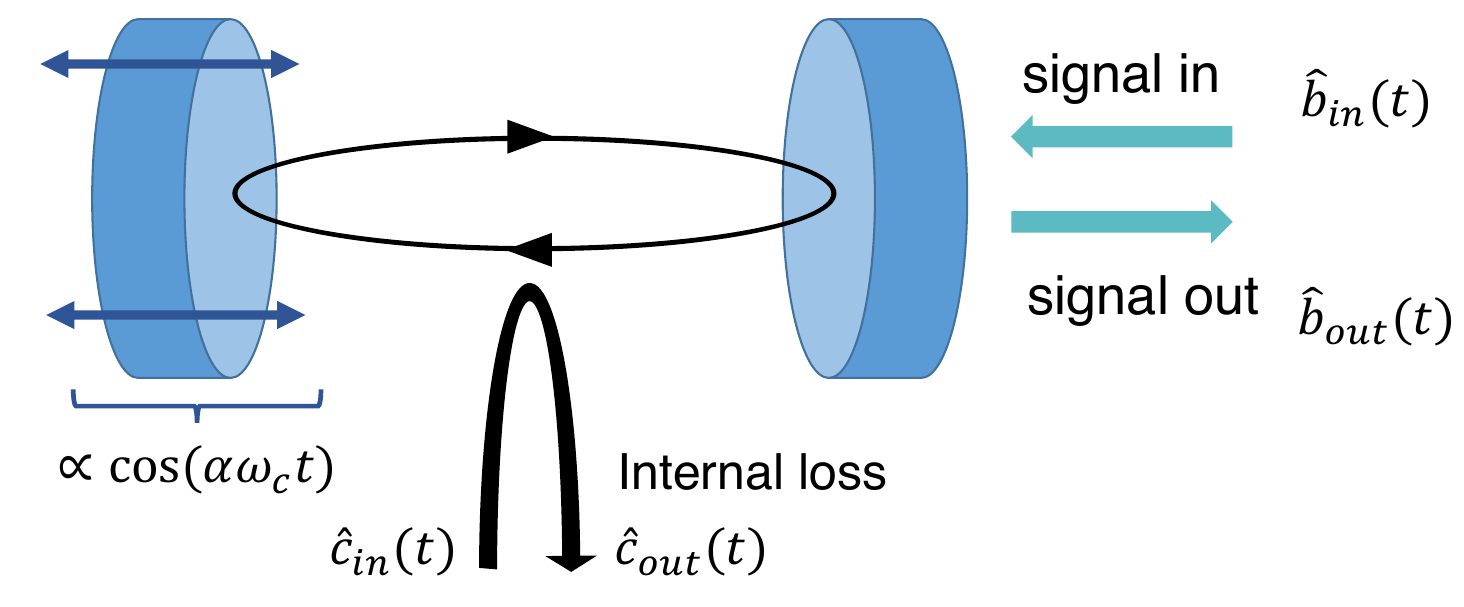}
    \caption{Schematic of the parametric amplifier. Here we use opto-mechanical analogy (resonator consisting of the cavity) instead of Josephson parametric amplifier.}
    \label{fig:optomechanics}
\end{figure}

We assume the resonator has a semi-infinite waveguide mode (the annihilation operator of which is denoted as $\hat{b}_{k}$) connected as an input/output port and also has internal losses in the resonator (the annihilation operator of which is denoted as $\hat{c}_{k}$). 
The schematic of this parametric amplifier using opto-mechanical analogy is \cref{fig:optomechanics}.
The total Hamiltonian describing this is 
\begin{align}
        H_{\text{tot}} &= H_{\text{sys}} +H_{\text{sig}}+H_{\text{loss}}, \\
        H_{\text{sys}} &=\omega_{c}\hat{a}^{\dagger}\hat{a}+
    \frac{\mu_{r}}{2}e^{i\alpha\omega_{c}t}\hat{a}^{2}+
    \frac{\mu_{r}}{2}e^{-i\alpha\omega_{c}t}\hat{a}^{\dagger 2}, \\
        H_{\text{sig}}&=\int \mathrm{d}\omega \left[\omega \hat{b}^{\dagger}(\omega) \hat{b}(\omega)
        +i \sqrt{\frac{\kappa_{e}}{2\pi}}(\hat{a}^{\dagger}\hat{b}(\omega)
        - \hat{b}^{\dagger}(\omega)\hat{a})
        \right], \\
        H_{\text{loss}}&=\int \mathrm{d}\omega \left[\omega \hat{c}^{\dagger}(\omega) \hat{c}(\omega)
        +i\sqrt{\frac{\kappa_{i}}{2\pi}}(\hat{a}^{\dagger}\hat{c}(\omega)
        - \hat{c}^{\dagger}(\omega)\hat{a})
        \right].
    \end{align}
Here, $\kappa_{e}$ is the external loss rate of the resonator, and $\kappa_{i}$ is the internal loss rate of the resonator.
As we did in \cref{sec:SNR}, we get Heisenberg equations for the resonator mode $\hat{a}$ and the input-output relationship of the waveguide:
\begin{align}
    \dv{\hat{a}(t)}{t} &= \left(-i\omega_{c} - \frac{\kappa}{2} \right)\hat{a}(t) -i\mu_{r}e^{-i\alpha \omega_{c}t}\hat{a}^{\dagger}(t) 
    + \sqrt{\kappa_{e}}\hat{b}_{\text{in}}(t)
    + \sqrt{\kappa_{i}}\hat{c}_{\text{in}}(t), 
    \label{eq:resonator} \\
    \hat{b}_{\text{out}} (t) &=\hat{b}_{\text{in}}(t) - \sqrt{\kappa_{e}}\hat{a}(t),
\end{align}
where $\kappa \equiv \kappa_{i}+\kappa_{e}$.

\subsection{Resonator equation}
In this subsection, we neglect the internal loss ($\kappa=\kappa_{e}$) and switch to a frame rotating at the angular frequency $\alpha\omega_{c}/2$, and deﬁne the following operators:
\begin{align}
    \hat{A}(t) &= e^{i\frac{\alpha}{2}\omega_{c}t}\hat{a}(t),\\
    \hat{B}_{\text{in}\,(\text{out})} (t) &=e^{i\frac{\alpha}{2}\omega_{c}t}\hat{b}_{\text{in}\,(\text{out})}(t) .
\end{align}
Assuming $\alpha=2$ for simplicity, the resonator equation \eqref{eq:resonator} and the input-output relation become
\begin{align}
    \dv{\hat{A}(t)}{t} &= - \frac{\kappa}{2} \hat{A}(t) -i\mu_r \hat{A}^\dag (t) + \sqrt{\kappa} \hat{B}_\mathrm{in} (t), \notag \\
    \hat{B}_{\text{out}} (t) &= \hat{B}_{\text{in}}(t) - \sqrt{\kappa}\hat{A}(t),
    \label{eq:input_output_B}
\end{align}
We consider the case with monochromatic incident light, i.e.,
\begin{equation}
    \hat{B}_\mathrm{in}(t) = \hat{B}_\mathrm{in}(0) e^{-i\Delta\omega},
\end{equation}
where $\Delta\omega\equiv \omega-\omega_c$.
In this case, the stationary solution of $\hat{A}(t)$ has only two Fourier components $e^{\pm i\Delta\omega t}$.
The resonator equations for these components are
\begin{equation}
    -i\Delta\omega \mqty(\hat{A}(\Delta\omega) \\ \hat{A}^\dag (-\Delta\omega))
        = \mqty(-\kappa/2 & -i\mu_r \\ +i\mu_r & -\kappa/2) \mqty(\hat{A}(\Delta\omega) \\ \hat{A}^\dag (-\Delta\omega))
        + \sqrt{\kappa} \mqty(\hat{B}_\mathrm{in}(0)\\ 0),
\end{equation}
and
\begin{equation}
    +i\Delta\omega \mqty(\hat{A}(-\Delta\omega) \\ \hat{A}^\dag (\Delta\omega))
        = \mqty(-\kappa/2 & -i\mu_r \\ +i\mu_r & -\kappa/2) \mqty(\hat{A}(-\Delta\omega) \\ \hat{A}^\dag (\Delta\omega))
        + \sqrt{\kappa} \mqty(0 \\ \hat{B}^\dag_\mathrm{in}(0)).
\end{equation}
Solving these equations, we obtain
\begin{align}
    \hat{A}(t) = \frac{\frac{\kappa}{2}-i\Delta\omega}{\qty(\frac{\kappa}{2}-i\Delta\omega)^2 - \mu_r^2} \sqrt{\kappa} \hat{B}_\mathrm{in}(0) e^{-i\Delta\omega t} + \frac{-i\mu_r}{\qty(\frac{\kappa}{2}+i\Delta\omega)^2 - \mu_r^2} \sqrt{\kappa} \hat{B}^\dag_\mathrm{in} (0) e^{+i\Delta\omega t}.
\end{align}
The output field is derived using \cref{eq:input_output_B} as
\begin{align}
    \hat{B}_\mathrm{out}(t) = & \qty[1 - \frac{\qty(\frac{\kappa}{2}-i\Delta\omega) \kappa}{\qty(\frac{\kappa}{2}-i\Delta\omega)^2 - \mu_r^2}] \hat{B}_\mathrm{in}(0) e^{-i\Delta\omega t} \notag \\
    & \qquad + \frac{-i\mu_r \kappa}{\qty(\frac{\kappa}{2}+i\Delta\omega)^2 - \mu_r^2} \hat{B}^\dag_\mathrm{in} (0) e^{+i\Delta\omega t}.
    \label{eq:B_out}
\end{align}
The first term represents the signal component, and the second term represents the idler component.

When $\Delta\omega=0$, these two modes degenerate. 
In this case, the output gain shows the phase-sensitivity.
In order to verify this, we define the following quadratures:

\begin{align}
    \hat{X}_{\theta} &\equiv \frac{\hat{B} e^{-i\theta}+\hat{B}^{\dagger} e^{i\theta}}{\sqrt{2}},\\
    \hat{Y}_{\theta} &\equiv \frac{\hat{B} e^{-i\theta}-\hat{B}^{\dagger} e^{i\theta}}{\sqrt{2}i}.
\end{align}
From \cref{eq:B_out} with $\Delta\omega=0$, we find
\begin{align}
    \hat{X}_{\theta,\,\text{out}}
        &=\left[1-
        \frac{\frac{\kappa^2}{2}}{\frac{\kappa^2}{4} - \mu_r^2}
        - \frac{\mu_r \kappa \sin(2\theta)}{\frac{\kappa^2}{4} - \mu_r^2}
        \right] \hat{X}_{\theta,\,\text{in}}
        - \frac{\mu_r \kappa \cos(2\theta)}{\frac{\kappa^2}{4} - \mu_r^2}
        \hat{Y}_{\theta,\,\text{in}}, 
        \label{eq:x_theta_measurement} \\
    \hat{Y}_{\theta,\,\text{out}}
        &=\left[1-
        \frac{\frac{\kappa^2}{2}}{\frac{\kappa^2}{4} - \mu_r^2}
        + \frac{\mu_r \kappa \sin(2\theta)}{\frac{\kappa^2}{4} - \mu_r^2}
        \right] \hat{Y}_{\theta,\,\text{in}}
        - \frac{\mu_r \kappa \cos(2\theta)}{\frac{\kappa^2}{4} - \mu_r^2}
        \hat{X}_{\theta,\,\text{in}}.
\end{align}
When $\theta = (1/4 + n)\pi$ ($n\in \mathbb{Z}$) in particular, they take the following form:
\begin{align}
    \hat{X}_{\theta,\text{out}}=\sqrt{G} \hat{X}_{\theta,\text{in}},
    \quad 
    \hat{Y}_{\theta,\text{out}}=\frac{1}{\sqrt{G}} \hat{Y}_{\theta,\text{in}},
    \label{eq:gain_x}
\end{align}
where the parameter $G$ is
\begin{equation}
    G = \qty(\frac{\mu_r + \frac{\kappa}{2}}{\mu_r - \frac{\kappa}{2}})^2 .
\end{equation}
\Cref{eq:gain_x} represents the squeezing by a JPA and is what we used in refs.~\cref{eq:squeeze_SQ,eq:squeeze_AMP}.

\bibliographystyle{jhep}
\bibliography{main}

\end{document}